\documentclass[prb,preprint,showpacs]{revtex4}
\usepackage{acronym}
\usepackage{amsfonts}
\usepackage{amsmath}
\usepackage{amssymb}
\usepackage{exscale}
\usepackage{graphicx}

\renewcommand{\mathbf}{\boldsymbol}

\begin{document}

\title{Non-collinear magnetic structures: a possible cause for current
  induced switching} 
\author{P. Weinberger$^{1)}$, A. Vernes$^{1)}$,
  B. L. Gy\"{o}rffy$^{1,2)}$, and L. Szunyogh$^{1,3)}$ }
\affiliation{$^{1)}$ Center for Computational Materials Science,
  Gumpendorferstr. 1a,
  A-1060 Vienna, Austria\\
  $^{2)}$ H. H. Wills Physics Laboratory, Bristol University, Royal
  Fort,
  Tyndall Avenue, Bristol BS8 1TL, U. K.\\
  $^{3)}$ Department of Theoretical Physics, Center for Applied
  Mathematics and Computational Physics, Budapest University of
  Technology and Economics, Budafoki \'{u}t. 8, H-1521 Budapest,
  Hungary} \date{April 21, 2004}

\begin{abstract}
  Current induced switching in Co/Cu/Co trilayers is described in
  terms of ab-initio determined magnetic twisting energies and
  corresponding sheet resistances. In viewing the twisting energy as
  an energy flux the characteristic time thereof is evaluated by means
  of the Landau-Lifshitz-Gilbert equation using ab-initio parameters.
  The obtained switching times are in very good agreement with
  available experimental data.  In terms of the calculated currents,
  scalar quantities since a classical Ohm's law is applied, critical
  currents needed to switch magnetic configurations from parallel to
  antiparallel and vice versa can unambiguously be defined. It is
  found that the magnetoresistance viewed as a function of the current
  is essentially determined by the twisting energy as a function of
  the relative angle between the orientations of the magnetization in
  the magnetic slabs, which in turn can also explain in particular
  cases the fact that after having switched off the current the system
  remains in the switched magnetic configuration. For all ab-initio
  type calculations the fully relativistic Screened
  Korringa-Kohn-Rostoker method and the corresponding Kubo-Greenwood
  equation in the context of density functional theory are applied.
\end{abstract}

\pacs{75.30.Gw, 75.70.Ak, 75.70.Cn}
\maketitle

\section{Introduction}

Reversal of the orientation of the magnetization without applying an
external field seems to be of considerable interest for magnetic
switching of micro-devices and caused extensive experimental and
theoretical studies of the effect of currents on magnetic
nanostructures. The experimental facts \cite%
{ex-11,ex-9,ex-15,ex-13,KAB+00,ex-6,AEM+02,ex-3,ex-5,ex-10,ex-20,GCJ+03,OKM+03}
are still quite confusing, or, to put in the words of a recent short
review article \cite{ex-1}, "many observed phenomena can be described
qualitatively ... by a simple semi-classical spin-torque model.
However, evidence of complications from several experiments suggests
that a full understanding of all observations is not yet achieved".
The by now generally accepted experimental facts are the following
ones: (1) if the current of a given sign favors the parallel (P)
magnetic configuration, the current of the opposite sign favors the
antiparallel (AP) configuration, (2) the current needed to switch the
magnetic configuration in nanostructured magnetic multilayer systems
is of the order of 2 - 5 mA in samples with a volume in the range of
40 - 800 nm$^{3}$. These two facts led \textit{inter alia }to a
schematic effective two-level energy diagram for switching in which
the critical current corresponds to the energy needed to overcome the
potential barrier between the parallel and the antiparallel magnetic
configuration.  Since experimentally also telegraph noise is observed,
which in turn seems to correspond to an oscillation between these two
states, this schematic picture proved to be quite useful. If by means
of a sufficiently high current the system is driven from one
configuration to the other one, it also can happen that after turning
off the current the system remains in the switched configuration,
i.e., the system does not return to the ground state. In the two-level
energy diagram this would correspond to the case that the two
schematic minima are separated by a high enough potential barrier and
are of about the same energy.

Most theoretical investigations \cite%
{th-8,th-10,BJZ98,th-9,th-6,th-7,th-12,th-2,th-4,th-1,th-5} were
concerned about finding expressions for the interaction between the
applied current and the orientation of the magnetic moments. Almost
all theoretical considerations and models used the concept of spin
currents and had to use phenomenological parameters to relate the
respective approach to the experimental evidence. Quite clearly in
most cases the main idea was to describe the cause for current induced
switching and deal afterwards with the subsequent effect, namely a
change in the magnetoresistance. Therefore the effect -- creating
excited states -- was interpreted in various ways by invoking spin
waves, all kinds of spin-polarization effects, etc. It is beyond the
scope of the present paper to summarize the various theoretical
approaches used up-to-now.

In here a completely different approach is pursued: the main idea is
to calculate fully relativistically the twisting (exchange
interaction) energy of a system as it goes from a parallel to an
antiparallel configuration, or opposite. This is a continuous function
of the relative angle between the orientations of the magnetization in
the magnetic parts of a spin valve system. In keeping one orientation
fixed and rotating the other one by an angle $\Theta$ around an axis
perpendicular to the fixed orientation one thus can switch
continuously from say the parallel magnetic configuration to the
antiparallel configuration. For each given rotation angle $\Theta$
simultaneously the corresponding sheet resistance (resistance divided
by the unit area) is calculated fully relativistically, which then is
also a continuous function of the rotation angle. It should be noted
that by using a fully relativistic approach the spin no longer is \ an
observable, i.e., at a given angle $\Theta$ there is just one sheet
resistance. In adopting this approach (1) the excitation energy is
related to the rotation angle, and (2) for the same angle a physical
observable, namely the sheet resistance is evaluated. Therefore at a
given $\Theta$ the effect of the physical phenomenon is described,
which then can be related to the cause, namely the turning on of a
current. It will be shown later on that the twisting (exchange
interaction) energy is the ab-initio analogon of the above mentioned
two-level energy diagram for switching. Furthermore, by means of
relating the twisting energy and the corresponding sheet resistance to
the current not only a critical current can be defined unambiguously,
but also the complexity of the switching process becomes evident.
Quite clearly in this picture no dynamic effects can be calculated,
although very good reasons for the occurrence of the telegraph noise
can be given. The quantum mechanical tools applied are the fully
relativistic screened Korringa-Kohn-Rostoker method \cite{th-15} and
the fully relativistic Kubo-Greenwood equation \cite{th-14} in the
context of the local density functional approximation. All further
reasoning is based on the Landau-Lifshitz-Gilbert equation, in terms
of which switching times can be evaluated using ab-initio parameters.
The introduced approach is applied to Co/Cu/Co type spin valves and in
fact will show quite a few of the experimentally observed features
mentioned earlier.

\section{Twisting energies and sheet resistances}

Consider a typical trilayer system of the type FM/NM$_{n}$/FM
consisting of two semi-infinite magnetic leads (FM) and a so-called
non-magnetic spacer (NM) such as for example Co(100)/Cu$_{n}$/Co(100)
or equivalently Co(100)/Cu$_{n}$/Co$_{m}$/Vac, where $n$ denotes the
number of spacer layers and $m$ is a sufficiently large number of
layers of the magnetic metal. Suppose now that
$\overrightarrow{n}_{0}$ denotes a particular unit vector (reference
orientation, either parallel or perpendicular to the surface normal)
characterizing the orientation of the magnetization in a particular
atomic layer containing one atom per unit cell. If
\begin{equation}
\overrightarrow{n}_{B}=\overrightarrow{n}_{B^{\prime}}=\overrightarrow{n}%
_{i}=\overrightarrow{n}_{0}\quad,\quad i=1,\ldots,n\ ,  \label{e2}
\end{equation}
such a configuration is usually referred to as a parallel configuration,
whereas for 
\begin{equation}
\overrightarrow{n}_{B}=\overrightarrow{n}_{i}=\overrightarrow{n}_{0}\ ,\quad
i=1,\ldots,\left( n/2\right) \ ;\quad\overrightarrow{n}_{B^{\prime}}=%
\overrightarrow{n}_{i}=-\ \overrightarrow{n}_{0}\ ,\quad i=(n/2)+1,\ldots
,n\ ,  \label{e3}
\end{equation}
frequently the term ``symmetric'' antiparallel configuration is used. If
$\overrightarrow{n}_{B}$, $\overrightarrow{n}_{B^{\prime}}$ (the
orientations of the magnetization in the semi-infinite leads) and the
$\overrightarrow {n} _{i}$ are each rotated by individual angles
around an axis perpendicular to $\overrightarrow{n}_{0}$ this
situation refers to a general non-collinear magnetic configuration in
two-dimensional translational invariant systems.  As for reasonably
large $n$ the interior of the NM part is completely non-magnetic in
the following specific non-collinear configurations of the type
\begin{equation}
\overrightarrow{n}_{B}=\overrightarrow{n}_{i}=\overrightarrow{n}%
_{0}\quad,\quad i=1,\ldots,\left( n/2\right) \ ;\quad\overrightarrow {n}%
_{B^{\prime}}=\overrightarrow{n}_{i}=\overrightarrow{n}_{0}^{\prime }\
,\quad i=(n/2)+1,\ldots,n\ ,  \label{e4}
\end{equation}
will be considered, where $\overrightarrow{n}_{0}^{\prime}$ is a unit vector
rotated by an angle $\Theta$ with respect to $\overrightarrow{n}_{0}$. It is
obvious that for these magnetic configurations it is sufficient to specify
the rotation angle $\Theta$. Expressed in simple terms this means that in
the right half of the trilayer system the orientation of the magnetization
is rotated uniformly by an angle $\Theta$ with respect to the orientation of
the magnetization in the left half.

Since a current perpendicular to the planes of atoms has to be described in
the present paper the reference orientation $\overrightarrow{n}_{0}$ is
chosen to be parallel to the surface normal ($z$ axis); the rotations are
performed around the $y$ axis.

\subsection{Twisting energies}

The energy difference between the two possible collinear states, namely the
parallel and the ``symmetric'' antiparallel magnetic configuration, is
usually termed interlayer exchange coupling energy. In using the magnetic
force theorem \cite{jansen} the total energies of these states are replaced
by the corresponding grand potentials (at zero temperature), i.e., by the
so-called band energy difference $\Delta E_{b}^{0}(\pi),$

\begin{equation}
\Delta E_{b}^{0}(\pi)=E_{b}(\pi)-E_{b}(0)=E_{b}(AP)-E_{b}(P)\ ,  \label{e5}
\end{equation}%
\begin{equation}
E_{b}(\Theta)=\int_{E_{0}}^{E_{F}}n(\Theta;E)(E-E_{F})dE\ ,  \label{e5-b}
\end{equation}
where $n(\Theta;E)$ is the density of states for a particular configuration, 
$E_{0}$ the valence band bottom and $E_{F}$ the Fermi energy. According to
Eq.~(\ref{e5}) the below convention applies 
\begin{equation}
\Delta E_{b}^{0}(\pi)=\left\{ 
\begin{array}{ccc}
>0 &  & P:\text{ground state} \\ 
<0 &  & AP:\text{ground state}%
\end{array}
\right. \ .  \label{e6}
\end{equation}
In a similar manner the ``twisting energy'' is defined by the following
difference, 
\begin{equation}
\Delta E_{b}(\Theta)=E_{b}(\Theta)-E_{b}(0)\ ,\quad0\leq\Theta\leq \pi\ ,
\label{e7}
\end{equation}
and -- as is well-known -- can be expanded in a power series in $\cos\Theta$%
\begin{equation}
\Delta E_{b}(\Theta)=a(1-\cos(\Theta))+b\cos^{2}(\Theta)+c\cos^{3}(\Theta)+\
\ldots\ ,  \label{e8}
\end{equation}
such that in all orders\textbf{\ }%
\begin{equation}
a=\Delta E_{b}(\pi/2)\ .  \label{e9}
\end{equation}
In first order $\Delta E_{b}(\Theta)$\textbf{\ } is then approximated by%
\begin{equation}
\Delta E_{b}(\Theta)\sim\Delta E_{b}^{(1)}(\Theta)=a(1-\cos(\Theta ))\ ,
\label{e10}
\end{equation}
in second order by%
\begin{equation}
\Delta E_{b}(\Theta)\sim\Delta E_{b}^{(2)}(\Theta)=a(1-\cos(\Theta))+b\cos
^{2}(\Theta)\ ,\quad b=\Delta E_{b}(\pi)-2\Delta E_{b}(\pi/2)\ ,  \label{e11}
\end{equation}
in third order by%
\begin{equation}
\Delta E_{b}(\Theta)\sim\Delta E_{b}^{(3)}(\Theta)=a(1-\cos(\Theta))+b\cos
^{2}(\Theta)+c\cos^{3}(\Theta)\ ,  \label{e13}
\end{equation}%
\begin{equation}
b=-\ \Delta E_{b}(\pi)-2\Delta E_{b}(\pi/2)+8\Delta E_{b}(2\pi/3)\ ,\quad
c=8\Delta E_{b}(2\pi/3)-2\Delta E_{b}(\pi)\ ,  \label{e14}
\end{equation}
\textbf{\ }etc., where $\Delta E_{b}(\pi),$ $\Delta E_{b}(\pi/2),\ldots,$
refer to the actually calculated values.

It should be noted that most frequently $\Delta E_{b}(\Theta)\sim\Delta
E_{b}^{(1)}(\Theta)$ is assumed, an approximation, which -- as will be shown
later on -- not necessarily is granted. Clearly enough in using only $\Delta
E_{b}^{(1)}(\Theta)$ the coefficient $a$ is simply half of the interlayer
exchange coupling energy, $a=\Delta E_{b}^{(1)}(\pi)/2$. In principle by
calculating $\Delta E_{b}(\Theta)$ for a few selected values of $\Theta,$ in
terms of Eq.\ (\ref{e8}) a reasonably good approximation to $\Delta
E_{b}(\Theta)$ for $\Theta$ varying continuously between $0$ and $\pi$ can
be obtained.

\subsection{Sheet resistance and magnetoresistance}

As is well-known in CPP (current perpendicular to the planes of atoms
geometry) the \ magnetoresistance can be defined via the sheet resistances
for the respective collinear magnetic configurations $P$ and $AP$,%
\begin{equation}
MR(\pi)=\frac{\Delta r(\pi)}{r(\pi)}=\frac{r(\pi)-r(0)}{r(\pi)}=\frac {%
r(AP)-r(P)}{r(AP)}\ ,  \label{e19}
\end{equation}
since the resistance $R(\Theta)$ is defined as 
\begin{equation}
R(\Theta)=r(\Theta)/A_{0}\ ,  \label{e19a}
\end{equation}
where $A_{0}$ is the unit area. In a similar manner for the present
non-collinear configurations, the difference in sheet resistances is given by%
\begin{equation}
\Delta r(\Theta)=r(\Theta)-r(0)\ ,  \label{e20}
\end{equation}
and the corresponding magnetoresistance by%
\begin{equation}
MR(\Theta)=\frac{\Delta r(\Theta)}{r(\Theta)}\ .  \label{e21}
\end{equation}
The difference in sheet resistances can again be expanded in a power series
in $\cos\Theta$%
\begin{equation}
\Delta r(\Theta)=\alpha(1-\cos(\Theta))+\beta\cos^{2}(\Theta)+\gamma\cos
^{3}(\Theta)+\ \ldots\ ,  \label{e22}
\end{equation}
with $\alpha=r(\pi/2)-r(0)$. It will be shown later on that in most of the
cases investigated in here 
\begin{equation}
\Delta r(\Theta)\sim\Delta r^{(1)}(\Theta)=\alpha(1-\cos(\Theta))\ ,
\end{equation}
i.e.,%
\begin{equation}
r(\Theta)=r(0)+\alpha(1-\cos(\Theta))\ .
\end{equation}

\subsection{Magnetic Joule's heat generated by a current}

For $I=0$, $\Theta$\ takes on its equilibrium value $\Theta_{\mathrm{eq}}$.
As the current $I$ is turned on, the relative orientation of the two
magnetic layers changes to $\Theta$. Evidently, the work done to accomplish
this rotation is $\Delta E(\Theta)=E(\Theta)-E(\Theta_{\mathrm{eq}})$.
Suppose that this energy difference is equal to the energy lost by the
current in the form of a ``magnetic'' contribution to the Joule's heat $Q$,%
\begin{equation*}
Q=R(\Theta)I^{2}\ .
\end{equation*}
Thus for a fixed current $I$,%
\begin{equation*}
\Delta E(\Theta)=\tau R(\Theta)I^{2}\ ,
\end{equation*}
where $\tau$\ is the time required to accomplish the rotation. This equation
can be solved for the function $\Theta\left( I\right) $, whose inverse is
given by
\begin{equation}
I(\Theta)=\pm\sqrt{A_{0}/\tau}\sqrt{r(\Theta)^{-1}\Delta E(\Theta )}\ ,
\label{e31}
\end{equation}
where%
\begin{equation}
\Delta E(\Theta)=E(\Theta)-E(0)+\min\left[ \Delta E(\Theta)\right] \ ,
\label{e33}
\end{equation}
i.e, where $\Delta E(\Theta)$ is a positive definite excitation energy.

Since $Q$, $\Delta E(\Theta)$ and therefore $I(\Theta)$ are scalar
positive definite quantities, the above construction is independent of the
direction of the current flow. Nevertheless, in the following, the
concept of twisting energies $\Delta E(\Theta)$ and the corresponding
magnetic Joule's heat generated during a time interval $\tau$ shall be used
to explore the physics of current induced switching. In short, evidence will 
be provided that the origin of the work done against the exchange forces 
acting between the two magnetic layers is the magnetic contribution to the 
energy dissipation from the current.

It might seem that by using an energy flux relation the problem of
evaluating the current $I$ was only shifted to yet another unknown quantity,
namely to the characteristic time $\tau$, whose theoretical description and
evaluation therefore has to be the subject of the next few sections.
Furthermore, it has to be pointed out that any comparison with experimental
data has to take into account also the actual area $A_{0}$ present in a
given experiment. However, before going ahead to discuss these two
quantities, the computational details of the ab-initio related parts of this
paper shall be given.

\subsection{Computational details}

The effective scattering potentials and exchange fields of spin valve
systems of the type fcc-Co(100)/Co$_{12}$/Cu$_{n}$/Co$_{m}$/Co(100), $12\leq
n\leq36$, $m\geq11$ were determined selfconsistently using the fully
relativistic spin-polarized Screened Korringa-Kohn-Rostoker method, \cite%
{th-15} where at least $m$ \ layers of Co served as "buffer" to the
semi-infinite leads. It should be noted that because of the special features
of the applied screened structure constants \cite{th-15} the total number of
atomic layers between the two semi-infinite systems has to be a multiple of
three. For this reason the thickness of the right buffer had to be kept
variable. In all cases the local density approximation of Vosko et al.\ \cite%
{Vosko} and, in order to obtain selfconsistency, a total of 45 $%
k_{\parallel} $ points in the irreducible part of the surface Brillouin zone
(IBZ) was applied. All selfconsistent calculations were performed with the
orientation of the magnetization pointing uniformly perpendicular to the
planes of atoms (reference configuration).

In using the magnetic force theorem the twisting energies with respect to
this reference configuration were then evaluated for each $n$ using the
symmetric arrangement, i.e., for the left half of the system (fcc-Co(100)/Co$%
_{12}$/Cu$_{n/2}$) the orientation of the magnetization remained unchanged
whereas the right half was rotated by a particular (uniform) angle $\Theta$.
For this kind of calculation a total of 960 $k_{\parallel}$ points in the
IBZ was used, a set-up, which yields very reliable results. For further
computational details concerning the Screened Korringa-Kohn-Rostoker method
and the evaluation of band energies, see the review article in Ref. %
\onlinecite{th-15}.

The sheet resistances for a given rotation angle $\Theta$ were first
evaluated by means of the fully relativistic version of the Kubo-Greenwood
equation\ at $E_{F}+i\delta,\delta>0$, and then numerically continued to the
real energy axis. In this part of the calculations for the occurring
Brillouin zone integrals a total of 1830 $k_{\parallel}$ points was used.
For a detailed discussion of this approach, see the review article in Ref. %
\onlinecite{th-14}.

\section{The Landau--Lifshitz--Gilbert equation}

From the polar form of the Landau-Lifshitz-Gilbert (LLG) equation, \cite%
{TBB02}%
\begin{equation}
\dfrac{d\mathcal{\vec{M}}}{dt}=\dfrac{\gamma_{\mathrm{G}}}{1+\alpha _{%
\mathrm{G}}^{2}}\left[ -\mathcal{\vec{M}}\times\vec{H}^{\mathrm{eff}}+\dfrac{%
\alpha_{\mathrm{G}}}{\mathcal{M}_{0}}\mathcal{\vec{M}}\times\left( \mathcal{%
\vec{M}}\times\vec{H}^{\mathrm{eff}}\right) \right] \ ,\qquad \left( \gamma_{%
\mathrm{G}},\alpha_{\mathrm{G}}>0\right) \ ,  \label{eq:Geq-polar}
\end{equation}
where $\gamma_{\mathrm{G}}$ is the Gilbert gyromagnetic ratio (precession
constant), $\alpha_{\mathrm{G}}$ is the dimensionless Gilbert damping
parameter, $\mathcal{\vec{M}}$ is the magnetization, $\mathcal{M}%
_{0}=\left\vert \mathcal{\vec{M}}\right\vert $, and $\vec{H}^{\mathrm{eff}}$
is the local and time--dependent effective field, one immediately observes
that (1) $\mathcal{\vec{M}}$ precesses almost purely, if damping is low ($%
\alpha_{\mathrm{G}}\rightarrow0$), \cite{LL99a} \ (2) almost no precession,
but slow switching occurs, when the damping is high ($\alpha_{\mathrm{G}%
}\rightarrow\infty$) \cite{Mal87}, and (3) the fastest switching refers to $%
\alpha_{\mathrm{G}}=1$. \cite{Mal00} Rewriting the LLG equation in terms of
an experimental\ damping parameter $G$, \cite{HUW02}%
\begin{equation}
\dfrac{1}{\gamma_{\mathrm{G}}}\dfrac{d\mathcal{\vec{M}}}{dt}=-\mathcal{\vec {%
M}}\times\vec{H}^{\mathrm{eff}}+\dfrac{G}{\gamma_{\mathrm{G}}^{2}\mathcal{M}%
_{0}^{2}}\left( \mathcal{\vec{M}}\times\dfrac{d\mathcal{\vec{M}}}{dt}\right)
\ ,  \label{eq:Geq-inialt}
\end{equation}
the dimensionless Gilbert parameter $\alpha_{\mathrm{G}}$ is given by \cite%
{HWU+03a} 
\begin{equation}
\alpha_{\mathrm{G}}=\dfrac{G}{\gamma_{\mathrm{G}}\mathcal{M}_{0}}\ ,
\label{eq:Geq-dimdamp}
\end{equation}
and the Gilbert gyromagnetic ratio $\gamma_{\mathrm{G}}$ by \cite%
{LL99a,BJZ98,GW01,SZ02} 
\begin{equation}
\gamma_{\mathrm{G}}=\dfrac{g\mu_{\mathrm{B}}}{\hbar}=\dfrac{g\left\vert
e\right\vert }{2m_{\mathrm{e}}}\ ,  \label{eq:Geq-gamma}
\end{equation}
where $e$ refers to the elementary charge, $m_{\mathrm{e}}$\ to the mass of
an electron, $\mu_{\mathrm{B}}$ to the Bohr magneton,\ and $g$ is the
(electronic) Land\'{e}--factor. Experimentally, \cite{UWH01,HWU+03a} it has
been shown that in multilayer systems $G$ in Eq.\ (\ref{eq:Geq-dimdamp})
varies linearly with $1\,/\,d$, where $d$ is the film thickness, see also
Tab.\ \ref{tab:expgilbertdamp}.

By definition the magnetization $\mathcal{\vec{M}}$ refers to the volume
averaged total magnetic moment. Assuming, however, that in a layered system
the layer--resolved magnetic moments $\vec{M}_{i}$, where $i$ denotes atomic
layers, are coherently precessing, \cite{Mil03} it is sufficient to describe
the magnetization dynamics of the layered system in terms of the motion of
either the layer averaged magnetic moment $\vec{M}$,%
\begin{equation}
\dfrac{d\vec{M}}{dt}=-\gamma\vec{M}\times\vec{H}^{\mathrm{eff}}+\alpha 
\dfrac{\vec{M}}{M_{0}}\times\left( \vec{M}\times\vec{H}^{\mathrm{eff}%
}\right) \ ,  \label{eq:LLGeq-avmom}
\end{equation}%
\begin{equation*}
\vec{M}=\frac{1}{N}\sum\limits_{i=1}^{N}\vec{M}_{i}\ ,
\end{equation*}
where $N$ denotes the number of magnetic layers, e.g.\ $N=m+n\,/\,2$, or in
terms of the magnetization direction $\overrightarrow{n}$ : \cite%
{TBB02,BNR03} 
\begin{equation}
\dfrac{d\overrightarrow{n}}{dt}=-\gamma\overrightarrow{n}\times\vec {H}^{%
\mathrm{eff}}+\alpha\overrightarrow{n}\times\left( \overrightarrow {n}\times%
\vec{H}^{\mathrm{eff}}\right) \ ,\qquad\ \overrightarrow{n}=\dfrac{\vec{M}}{%
M_{0}}\ .  \label{eq:LLGeq-norm}
\end{equation}
It should be noted that the equivalence of these two equations, namely Eqs. (%
\ref{eq:LLGeq-avmom}) and (\ref{eq:LLGeq-norm}), relies on the conservation
of $\mathcal{M}$ , which in turn implies that $\left\vert \vec{M}\right\vert
=M_{0}$ and $\left\vert \overrightarrow{n}\right\vert =1$.

\subsection{Internal effective field}

The local effective field $\vec{H}^{\mathrm{eff}}$ that enters Eqs.\ (\ref%
{eq:LLGeq-avmom}) and (\ref{eq:LLGeq-norm}) can directly be derived from the
Helmholtz free energy density by taking its variational derivative with
respect to the magnetization, \cite{Mal00,HWU+03a,GW01,KK02}%
\begin{equation}
\vec{H}^{\mathrm{eff}}=-\dfrac{\partial\mathcal{F}}{\partial\mathcal{\vec{M}}%
}=-\mathbf{\nabla}_{\mathcal{\vec{M}}}\mathcal{F}\ ,\qquad\mathrm{with}\quad%
\mathcal{F}=\dfrac{F}{V}  \label{eq:eff-field}
\end{equation}
where $V$ is the total characteristic volume of the system and the free
energy $F$ includes the exchange energy, the crystalline anisotropy energy,
external magnetic fields, etc., \cite{BNR03,APC00} either in a
parameter--free manner or by using different types of model Hamiltonians.
Since for layered systems as considered in here,%
\begin{equation*}
\mathbf{\nabla}_{\mathcal{\vec{M}}}=\Omega_{0}\sum_{\mu=\mathrm{x},\mathrm{y}%
,\mathrm{z}}\vec{e}_{\mu}\,\dfrac{\partial}{\partial M_{\mu}}=\Omega_{0}%
\mathbf{\nabla}_{\vec{M}}\ \ ,
\end{equation*}
Eq.\ (\ref{eq:eff-field}) can be written as 
\begin{equation}
\vec{H}^{\mathrm{eff}}=\dfrac{\partial\overline{F}}{\partial\vec{M}}=-%
\mathbf{\nabla}_{\vec{M}}\overline{F}\ ,  \label{eq:eff-field4ml}
\end{equation}
where the Helmholtz free energy $\overline{F}$,%
\begin{equation*}
\overline{F}=\dfrac{1}{N}\sum_{i=1}^{N}F_{i\ },
\end{equation*}
refers to the reference volume $\Omega_{0}$ and $N$ is again the number of
magnetic layers considered.

According to Eq.\ (\ref{eq:eff-field4ml}) the internal effective field, $%
\vec{H}^{\mathrm{E}}$, arises from the contribution of the total energy $%
E_{b}$\ to the free energy, 
\begin{equation*}
\vec{H}^{\mathrm{E}}=-\dfrac{\partial E_{b}}{\partial\vec{M}}=-\mathbf{%
\nabla }_{\vec{M}}E_{b}\ .
\end{equation*}
Assuming, e.g., in terms of Eq.(\ref{e8}) that the derivatives 
\begin{equation*}
\dfrac{\partial^{k}E_{b}(\vec{M}_{0})}{\partial M_{\mathrm{x}%
}^{k_{1}}\partial M_{\mathrm{y}}^{k_{2}}\partial M_{\mathrm{z}}^{k_{3}}}\
,\qquad\sum_{j=1}^{3}k_{j}=k=1,\ldots,n\quad(k_{j},\,k,\,n\in\mathbb{N})\ ,
\end{equation*}
in the below Taylor series expansion of the total energy,%
\begin{equation}
E_{b}(\vec{M}_{0}+\vec{M})\simeq E_{b}(\vec{M}_{0})+\sum_{k=1}^{p}\dfrac {1}{%
k!}\left( \vec{M}\cdot\mathbf{\nabla}_{\vec{M}}\right) ^{k}E_{b}(\vec {M}%
_{0})\ ,  \label{eq:totentay}
\end{equation}
\ are available up to a certain order $p$, where $\vec{M}_{0}$ is the
initial reference moment, the Cartesian components of the internal effective
field can directly be given.

In particular, if the change in the moment $\vec{M}$ is constrained to the $0%
\mathrm{yz}$-plane, 
\begin{equation*}
\vec{M}=M_{\mathrm{y}}\vec{e}_{\mathrm{y}}+M_{\mathrm{z}}\vec{e}_{\mathrm{z}%
}\ ,\qquad\left( M_{\mathrm{x}}=0\right) \ ,
\end{equation*}
then%
\begin{equation}
\vec{H}^{\mathrm{E}}=-\,\sum_{k=1}^{p}\sum_{q=0}^{k}\dfrac{\partial^{k}E(%
\vec{M}_{0})}{\partial M_{\mathrm{y}}^{k-q}\partial M_{\mathrm{z}}^{q}}\left[
\dfrac{M_{\mathrm{y}}^{k-q-1}M_{\mathrm{z}}^{q}}{\left( k-q-1\right) !q!}%
\vec{e}_{\mathrm{y}}\,+\,\dfrac{M_{\mathrm{y}}^{k-q}M_{\mathrm{z}}^{q-1}}{%
\left( k-q\right) !\left( q-1\right) !}\,\vec{e}_{\mathrm{z}}\right] \ ,
\label{eq:Heff4Epw}
\end{equation}
with $\vec{M}_{0}=M_{0}\vec{e}_{\mathrm{z}}$ being the initial, ground state
moment. Provided that the magnitude of the moment is preserved,%
\begin{equation}
M_{0}^{2}=M_{\mathrm{y}}^{2}+M_{\mathrm{z}}^{2}\ =M^{2},
\label{eq:magnconst}
\end{equation}
by keeping in Eq.\ (\ref{eq:totentay}) only terms up $p=3$, one gets 
\begin{equation}
\Delta E_{b}\left( \vec{M}\right) =E_{b}(\vec{M}_{0}+\vec{M})-E_{b}(\vec {M}%
_{0})\simeq a-a\dfrac{M_{\mathrm{z}}}{M_{0}}+b\,\dfrac{M_{\mathrm{z}}^{2}}{%
M_{0}^{2}}+c\dfrac{M_{\mathrm{z}}^{3}}{M_{0}^{3}}\ ,
\label{eq:totenpwtay3rd}
\end{equation}
where the coefficients $a$, $b$ and $c$ are defined in Eqs.\ (\ref{e9}) and (%
\ref{e14}). Therefore the energy torque rotating the moment is given by 
\begin{equation}
\vec{M}\times\vec{H}^{\mathrm{E}}=\vec{e}_{\mathrm{x}}M_{\mathrm{y}}H_{%
\mathrm{z}}^{\mathrm{E}}=-n_{\mathrm{y}}\left( -a+2b\,n_{\mathrm{z}}+3cn_{%
\mathrm{z}}^{2}\right) \,\vec{e}_{\mathrm{x}}\ ,  \label{eq:epw3rdtorque}
\end{equation}
whereas%
\begin{equation}
\vec{M}\times\left( \vec{M}\times\vec{H}^{\mathrm{E}}\right) =-n_{\mathrm{y}%
}\left( -a+2b\,n_{\mathrm{z}}+3cn_{\mathrm{z}}^{2}\right) \left( M_{\mathrm{z%
}}\,\vec{e}_{\mathrm{y}}-M_{\mathrm{y}}\,\vec{e}_{\mathrm{z}}\right) \ .
\label{eq:epw3rdtorquexm}
\end{equation}

\subsection{The characteristic time of switching}

Inserting the internal effective field $\vec{H}^{\mathrm{E}}$ into the LLG
equation (in the absence of precession around the $\mathrm{z}$-axis), 
\begin{equation*}
\dfrac{d\vec{M}}{dt}\simeq\alpha\dfrac{\vec{M}}{M_{0}}\times\left( \vec {M}%
\times\vec{H}^{\mathrm{E}}\right) \ ,\quad\alpha=\,\alpha_{\mathrm{G}}\dfrac{%
\gamma_{\mathrm{G}}}{1+\alpha_{\mathrm{G}}^{2}}\ ,
\end{equation*}
\ then leads to%
\begin{align}
M_{0}\dfrac{dn_{\mathrm{x}}}{dt} & =0\ ,  \label{eq:LLGeq-pw-1} \\
M_{0}\dfrac{dn_{\mathrm{y}}}{dt} & =-\alpha n_{\mathrm{y}}n_{\mathrm{z}%
}\left( -a+2b\,n_{\mathrm{z}}+3cn_{\mathrm{z}}^{2}\right) \ ,
\label{eq:LLGeq-pw-2} \\
M_{0}\dfrac{dn_{\mathrm{z}}}{dt} & =\alpha n_{\mathrm{y}}^{2}\left(
-a+2b\,n_{\mathrm{z}}+3cn_{\mathrm{z}}^{2}\right) \ .  \label{eq:LLGeq-pw-3}
\end{align}
Since according to Eq.\ (\ref{eq:magnconst}), $n_{\mathrm{y}}^{2}+n_{\mathrm{%
z}}^{2}\ =1$, Eq.\ (\ref{eq:LLGeq-pw-3}) reduces to%
\begin{equation}
M_{0}\dfrac{dn_{\mathrm{z}}}{dt}=\alpha\left( 1-n_{\mathrm{z}}^{2}\right)
\left( -a+2b\,n_{\mathrm{z}}+3cn_{\mathrm{z}}^{2}\right) \ .
\label{eq:mzeq-pw}
\end{equation}
Assuming that $n_{\mathrm{z}}\neq\pm1$ or$\ \left( -b\pm\sqrt{b^{2}+3ac}%
\right) \,/\,3c$ and $b^{2}+3ac>0$, $c\neq0$,\ Eq.\ (\ref{eq:mzeq-pw}) can
directly be integrated and leads to the time $\tau=t_{f}-t_{i}$ needed to
change $n_{\mathrm{z}}$ from $n_{\mathrm{z}}^{i}=n_{\mathrm{z}}\left(
t_{i}\right) $ to $n_{\mathrm{z}}^{f}=n_{\mathrm{z}}\left( t_{f}\right) $ 
\begin{align}
\dfrac{\alpha}{M_{0}}\tau & =\dfrac{1}{2\left[ \left( 3c-a\right) -2b\right] 
}\ln\left\vert \dfrac{n_{\mathrm{z}}^{f}+1}{n_{\mathrm{z}}^{i}+1}\right\vert
-\dfrac{1}{2\left[ \left( 3c-a\right) +2b\right] }\ln\left\vert \dfrac{n_{%
\mathrm{z}}^{f}-1}{n_{\mathrm{z}}^{i}-1}\right\vert  \notag \\
& -\dfrac{b}{\left( 3c-a\right) ^{2}-4b^{2}}\ln\left\vert \dfrac{3c\left( n_{%
\mathrm{z}}^{f}\right) ^{2}+2b\,n_{\mathrm{z}}^{f}-a}{3c\left( n_{\mathrm{z}%
}^{i}\right) ^{2}+2b\,n_{\mathrm{z}}^{i}-a}\right\vert  \notag \\
& +\dfrac{1}{\left( 3c-a\right) ^{2}-4b^{2}}\dfrac{a\left( 3c-a\right)
+2b^{2}}{2\sqrt{b^{2}+3ac}}  \notag \\
& \qquad\qquad\times\ln\left\vert \dfrac{\left( b+3cn_{\mathrm{z}%
}^{f}\right) -\sqrt{b^{2}+3ac}}{\left( b+3cn_{\mathrm{z}}^{i}\right) -\sqrt{%
b^{2}+3ac}}\dfrac{\left( b+3cn_{\mathrm{z}}^{i}\right) +\sqrt {b^{2}+3ac}}{%
\left( b+3cn_{\mathrm{z}}^{f}\right) +\sqrt{b^{2}+3ac}}\right\vert \ .
\label{eq:dt-pw}
\end{align}
Thus the time $\tau\left( \alpha_{\mathrm{G}}\right) $ needed to change $n_{%
\mathrm{z}}$ from $n_{\mathrm{z}}^{i}$ to $n_{\mathrm{z}}^{f}$ as a function
of the Gilbert damping parameter $\alpha_{\mathrm{G}}>0$, when using $%
\alpha=\,\alpha_{\mathrm{G}}\gamma_{\mathrm{G}}/\left( 1+\alpha_{\mathrm{G}%
}^{2}\right) $ and Eq.\ (\ref{eq:Geq-gamma}), can be written as 
\begin{equation}
\tau\left( \alpha_{\mathrm{G}}\right) =C\dfrac{M_{0}}{\gamma_{\mathrm{G}}}%
\dfrac{1+\alpha_{\mathrm{G}}^{2}}{\alpha_{\mathrm{G}}}=\hbar\dfrac {%
\mathfrak{M}_{0}}{g}\,C\,\dfrac{1+\alpha_{\mathrm{G}}^{2}}{\alpha _{\mathrm{G%
}}}\ ,  \label{eq:dtse-pw}
\end{equation}
where $C$ denotes the rhs of Eq.\ (\ref{eq:dt-pw}) and $\mathfrak{M}_{0}$ is
the magnetic moment $M_{0}$ in units of Bohr magnetons $\mu_{\mathrm{B}}$.

According to Eq.\ (\ref{eq:dtse-pw}) it follows that the minimal time of
changing $n_{\mathrm{z}}$ from $n_{\mathrm{z}}^{i}$ to $n_{\mathrm{z}}^{f}$
is given by 
\begin{equation}
\tau_{\mathrm{\min}}=\tau\left( \alpha_{\mathrm{G}}=1\right) =2\hbar \dfrac{%
\mathfrak{M}_{0}}{g}\,C\,\qquad\mathrm{and}\qquad\tau\left( \alpha_{\mathrm{G%
}}\right) =\dfrac{1+\alpha_{\mathrm{G}}^{2}}{\alpha _{\mathrm{G}}}\,\dfrac{%
\tau_{\mathrm{\min}}}{2}\ .  \label{eq:dtmin-pw}
\end{equation}

\section{Results}

\subsection{Magnetization and switching times}

The selfconsistently obtained constant magnitude of the magnetization for
fcc Co$/$Cu$/$Co(100) and in the case of a thin Co slab of $\mathcal{M}_{0}^{%
\mathrm{fcc\ Co}}=1.418\times10^{6}\ \mathrm{A\ m}^{-1}$ \ is in very good
agreement with the available experimental data for Co bulk, see Ref.\ %
\onlinecite{SRR03} or \onlinecite{MRK+99}. By using the experimental Land%
\'{e} g--factor for fcc Co, namely \ $g^{\mathrm{fcc\ Co}}=2.146\pm 0.02$,%
\cite{SPF+95} $\gamma_{\mathrm{G}}^{\mathrm{fcc\ Co}}=18.87213449\times
10^{10}\,\mathrm{m\ A}^{-1}\mathrm{\,s}^{-1}$, which in turn yields the
below ratio%
\begin{equation*}
\dfrac{M_{0}^{\mathrm{fcc\ Co}}}{\gamma_{\mathrm{G}}^{\mathrm{fcc\ Co}}}%
=\hbar\dfrac{\mathfrak{M}_{0}^{\mathrm{fcc\ Co}}\ }{g^{\mathrm{fcc\ Co}}}%
=0.782531193\times10^{-34}\ \mathrm{Js}=4.884173446\times10^{-16}\ \mathrm{%
eVs}\ .
\end{equation*}
Since according to Tab.\ \ref{tab:abc} the quantities on the rhs of Eq.\ (%
\ref{eq:dt-pw}) are of the order of $\left( \mathrm{meV}\right)
^{-1}=10^{3}\ \left( \mathrm{eV}\right) ^{-1}$, this implies that the time
needed to change the moment direction from $n_{\mathrm{z}}^{i}$ to $n_{%
\mathrm{z}}^{f}$ is of the order of $\ 10^{-13}\ \mathrm{s}$, namely
femtoseconds. From Tab.\ \ref{tab:taumin} one immediately can see that the
theoretically obtained values of $\tau_{\mathrm{\min}}$ are within the range
of values known from micromagnetic simulations for a polycrystalline thin Co
film, which showed that the reversal time ranges from $0.05$ $\mathrm{ns}$
for $\alpha_{\mathrm{G}}=1$ to $0.2$ $\mathrm{ns}$ for $\alpha_{\mathrm{G}%
}=0.1$ . \cite{SRR03} It should be noted that because the sign of $\tau_{%
\mathrm{\min}}$ is uniquely determined by the sign of the initial and final
values for $n_{\mathrm{z}}$, in Tab.\ \ref{tab:taumin} only those values for$%
\ n_{\mathrm{z}}^{i}$ and $n_{\mathrm{z}}^{f}$ are listed, which yield $%
\tau_{\mathrm{\min}}>0$. The sign of the such determined $n_{\mathrm{z}}^{f}$
\ confirms therefore independently the ground state configuration predicted
by the magnetic force theorem, see Eq.~(\ref{e7}).

As can be seen from the corresponding column in Tab.\ \ref{tab:taumin} the
switching time is largest for the Gilbert damping parameter of Co bulk. By
scaling $\alpha_{\mathrm{G}}$ to the thickness of the thin (rotated) Co slab
used in the present calculations according to values found for Co$_{N}/$Cu$/$%
Co(100), \cite{MRK+99} the magnetization reversal time $\tau\left( \alpha_{%
\mathrm{G}}\right) $ in Eq.~(\ref{eq:dtmin-pw}) changes only very moderately
in comparison with $\tau_{\mathrm{\min}}$.

\subsection{The importance of cross sections (unit areas)}

Going now back to Eq. (\ref{e31}), rewritten below by indicating the
appropriate units,

\begin{equation}
\left. I\left( \Theta\right) \right\vert _{\mathrm{SI}}=\pm1.265771437\cdot 
\sqrt{\dfrac{\left\langle A_{0}\right\rangle _{\mathrm{SI}}}{\left\langle
\tau\right\rangle _{\mathrm{SI}}}}\sqrt{\dfrac{\left\langle \Delta E\left(
\Theta\right) \right\rangle _{\mathrm{meV}}}{\left\langle r\left(
\Theta\right) \right\rangle _{\mathrm{m\Omega\cdot\mu m}^{2}}}}\ \mathrm{mA}%
\ ,  \label{e-final}
\end{equation}
it is obvious that for any kind of comparison to experiment not only $\tau$
has to be evaluated, but also that $A_{0}$, the surface\ perpendicular to
the $\mathrm{z}$ axis through which the current $I\left( \Theta\right) $
flows, has to be taken into account. Usually the cross section of
nanopillars is given in nanometer ($\mathrm{nm)}$, i.e., is of the order of 
\begin{equation*}
\left\langle A_{0}\right\rangle _{\mathrm{SI}}=\left\langle
A_{0}\right\rangle _{\mathrm{nm}^{2}}\times10^{-18}\ ,
\end{equation*}
which combined with the switching time (in nanoseconds),%
\begin{equation*}
\left\langle \tau\right\rangle _{\mathrm{SI}}=\left\langle \tau\right\rangle
_{\mathrm{ns}}\times10^{-9}\ ,
\end{equation*}
yields the following factor that multiplies the square root of the (quantum
mechanically derived) quotient of twisting energy and sheet resistance in
Eq. (\ref{e-final}),%
\begin{equation*}
\sqrt{\dfrac{\left\langle A_{0}\right\rangle _{\mathrm{SI}}}{\left\langle
\tau\right\rangle _{\mathrm{SI}}}}=\sqrt{\dfrac{\left\langle
A_{0}\right\rangle _{\mathrm{nm}^{2}}\times10^{-18}}{\left\langle
\tau\right\rangle _{\mathrm{ns}}\times10^{-9}}}=\sqrt{\dfrac{\left\langle
A_{0}\right\rangle _{\mathrm{nm}^{2}}}{\left\langle \tau\right\rangle _{%
\mathrm{ns}}}\times10^{-9}}\ .
\end{equation*}
In using, e.g., $\left\langle A_{0}\right\rangle _{\mathrm{nm}^{2}}=120000$ 
\cite{GCJ+03}, other values of $\left\langle A_{0}\right\rangle _{\mathrm{nm}%
^{2}}$ are listed in Tab.\ \ref{tab:cross-sect}, and $\left\langle
\tau_{\min }\right\rangle _{\mathrm{ns}}=0.01$, this results into a value
for $\sqrt{\left\langle A_{0}\right\rangle _{\mathrm{SI}}/\left\langle
\tau_{\min }\right\rangle _{\mathrm{SI}}}$ of about 0.11.

\subsection{Twisting energies and currents}

In all figures showing twisting energies and sheet resistances, etc., the
actually calculated values are displayed, solid lines only serve as guidance
to the eye. For illustrative purposes also the first order approximation to
the twisting energy is depicted in these figures as a dashed line. As it is
not possible to show all results obtained these figures concentrate on
systems with the spacer thickness varying between about 35 - 50~\AA . This
still results in a considerable number of figures, which, however, seems to
be necessary considering that in experimental studies mostly nanopillars are
used, i.e., most likely an average over thicknesses is recorded, and also in
order to illustrate the complexity of the effects to be seen. Furthermore,
in all figures for the current $I\left( \Theta\right) $ the factor $\sqrt{%
\left\langle A_{0}\right\rangle _{\mathrm{SI}}/\left\langle \tau_{\min
}\right\rangle _{\mathrm{SI}}}$ in Eq. (\ref{e-final}) is replaced by unity.

In the investigated range of spacer thicknesses the number of cases in which
the twisting energy is proportional to $(1-\cos(\Theta))$, see the
corresponding figures for $n\geq31$, is surprisingly small, whereas in all
cases the sheet resistance -- more or less -- is of this shape. This in turn
implies that all special features to be seen for $I(\Theta)$ are mostly
related to the functional form of the twisting energy. Taking for example $%
n=20$, $I(\Theta)$ remains about constant for $\Theta\geq90^{0}$, a value
which refers also to the critical current that has to be applied to drive
the system from parallel to antiparallel. However, one also can see from
this figure that $\Delta E_{b}(\Theta)$ has a maximum at about $140^{0}$:
the system has to overcome a small barrier to return to the ground state
(parallel configuration). From the entry showing the magnetoresistance
versus current, it is evident that at the critical current the
magnetoresistance jumps by about 20\%. For $n=21$ the situation is even more
dramatic, since $\Delta E_{b}(\Theta)$ has quite a large maximum at $90^{0}$%
, the AP configuration being only slightly less energetically favored than
the P configuration. In this particular system the meaning of the critical
current is quite obvious: it simply is the maximum in the $I(\Theta)$ versus 
$\Theta$ curve. The same situation, even more impressive, pertains for $n=25$%
, since now the parallel and antiparallel configuration are virtually
degenerated in energy, separated, however, by quite a barrier. The figures
for $n=20,21,23,24,25$ and $27$ are perfect ab-initio analoga for the
schematic effective two-level energy diagram mentioned in the introduction:
the energy displayed in this \textit{ad hoc} scheme is nothing but the
twisting energy, the schematic abscises being the relative angle between the
two orientations of the magnetization.

The system with 26 Cu spacer layers is in particular interesting, since a
non-collinear configuration is the ground state. In this case a tiny current
(about 0.05 mA in this figure) produces a magnetoresistance that can be
either zero, 1.5 or about 6 \%. It should be noted that the energy barrier
between the ground state and the parallel configuration is minute: the
system can almost freely oscillate between magnetic configurations for
values of $\Theta\leq60^{0}$. Another interesting case seems to be for 30 Cu
spacer layers, which shows a strong deviation from $\Delta
E_{b}^{(1)}(\Theta)$ at about $\Theta=90^{0}$, not enough, however, to cause
an additional minimum between the AP (ground state) and the P configuration.
For $0\leq\Theta \leq100^{0}$ the current $I(\Theta)$ varies almost linear,
changes slope, and varies again almost linear for $\Theta>100^{0}$.

For $n\geq31$ no more interesting effects are observed: the twisting energy
can be described very well in terms of \ $\Delta E_{b}^{(1)}(\Theta)$; in
order to switch from parallel to antiparallel or vice versa a current of
about 0.35 - 0.6 mA is needed. It should be recalled that all values of $%
I(\Theta)$ quoted in this section refer to $\sqrt{\left\langle
A_{0}\right\rangle _{\mathrm{SI}}/\left\langle \tau_{\min}\right\rangle _{%
\mathrm{SI}}}=1$ in Eq. (\ref{e-final}).

In the last figure finally the rotation of the magnetization around the
z-axis (precession) is shown for $n=25$. As can be seen the precessional
changes in the twisting energy are very small indeed. This figure justifies 
\textit{a posteri} the approach taken to evaluate and discuss the switching
time $\tau _{\min }$.

\section{Discussion}

In viewing now all the various cases discussed above the following
observations can be added: (1) if the slope of the magnetoresistance with
respect to the current is uniformly positive (negative), the parallel
(antiparallel) magnetic configuration is favoured (see, $n=31,33$ versus $%
n=32$), (2) if it becomes approximately infinite at a certain current then a
jump in the magnetoresistance occurs (e.g., $n=20,24,28$), and (3) if this
slope changes sign, a more complicated behavior pertains (e.g., $%
n=21,22,25,26$). In the latter case the system either remains in the
switched configuration ($n=21,25$) or because of a non-collinear ground
state oscillations in the magnetoresistance between zero and a few percent
can occur when a very small current is applied ($n=26$). The so-called
telegraph noise seems to refer to the jumping between such minima in the
twisting energy, the jumping rates obviously being connected with the
barrier between these minima. The current needed to switch a configuration
from parallel to antiparallel (or vice versa) refers to the largest value of 
$I(\Theta)$.

The present results suggest that the efficiency of current-driven switching
can considerably be optimized by varying the spacer thickness: theoretically
in using a spacer thickness of about 43~\AA\ (25 layers of Cu) perfect
switching can be achieved such that the system remains in the switched state
after the current is turned off. Further theoretical investigations using
the approach presented in here can include interdiffusion effects at
interfaces or refer to different kinds of magnetic slabs (leads) such as for
example permalloy, a system, which because of anisotropy effects in the
magnetoresistance perhaps is even more complicated than the present Co/Cu/Co
trilayer.

Altogether correlating the twisting energy and the corresponding resistance
with the current yields a very consistent view of the complexity found in
current-driven experiments. Clearly enough this correlation suffers from the
fact that up-to-now no quantum mechanical description for the Gilbert
damping factor was found and that a linear response theory (Kubo-Greenwood
equation) is used to evaluate the electric transport properties, i.e., that
the current had to be formulated as a scalar quantity.

Finally, it has to be remarked that the experimentally observed critical
switching currents are by a factor of about 10 - 100 larger than the ones
obtained in here. It has to be remembered, however, that in here an ideal
Co/Cu/Co trilayer was assumed while most experiments are based on rather
complicated nanostructures such as for example nanopillars and therefore -
although a consistent approach to was introduced in order to evaluate
switching times in terms of ab-initio parameters - also the question of a
comparable cross section (unit area) is of quite some importance.

\section{Acknowledgement}

The authors are grateful for many fruitful discussions with Profs. P.
M. Levy, J. Bass and C. Sommers. This work has been financed by the
Center for Computational Materials Science (Contract No. GZ 45.531), a
special grant from the Technical University of Vienna, and the
Hungarian National Scientific Research Foundation (Contracts OTKA
T037856 and OTKA T046267).

\pagebreak


\begin{center}
\begin{table}[hbtp] \centering%
\caption{\label{tab:expgilbertdamp} Experimental damping parameter
$\protect G$ for different systems. }%
\begin{tabular}{|c|c|c|}
\hline
\textbf{material} & \textbf{type of system} & $G$ $\left( 10^{8}\ \mathrm{s}%
^{-1}\right) $ \\ \hline
\multicolumn{1}{|l|}{Fe} & \multicolumn{1}{|l|}{bulk} & \multicolumn{1}{|l|}{%
$0.5$ \cite{HUW02}, $0.58$ \cite{HUA+87}} \\ \hline
\multicolumn{1}{|l|}{} & \multicolumn{1}{|l|}{} & \multicolumn{1}{|l|}{$%
0.59\pm0.06;\ 0.572\pm0.04$ \cite{SPF+95}} \\ \hline
\multicolumn{1}{|l|}{} & \multicolumn{1}{|l|}{} & \multicolumn{1}{|l|}{$0.8$ 
\cite{HUW02a} $0.7\pm0.06$ \cite{SPF+95}} \\ \hline
\multicolumn{1}{|l|}{Fe} & \multicolumn{1}{|l|}{single film} & 
\multicolumn{1}{|l|}{$1.5$ \cite{UWH01,HWU+03a}} \\ \hline
\multicolumn{1}{|l|}{} & \multicolumn{1}{|l|}{} & \multicolumn{1}{|l|}{$%
1.3\pm 0.1$ \cite{HUW02a}} \\ \hline
\multicolumn{1}{|l|}{Fe / Ag(100)} & \multicolumn{1}{|l|}{$d_{\mathrm{Fe}%
}=40\ \mathrm{\mathring{A}}$} & \multicolumn{1}{|l|}{$0.66$ \cite{HUA+87}}
\\ \hline
\multicolumn{1}{|l|}{} & \multicolumn{1}{|l|}{$d_{\mathrm{Fe}}=24\ \mathrm{%
\mathring{A}}$} & \multicolumn{1}{|l|}{$0.65$ \cite{HUA+87}} \\ \hline
\multicolumn{1}{|l|}{} & \multicolumn{1}{|l|}{$d_{\mathrm{Fe}}=7\ \mathrm{%
\mathring{A}}$} & \multicolumn{1}{|l|}{$2.3$ \cite{HUA+87}} \\ \hline
\multicolumn{1}{|l|}{} & \multicolumn{1}{|l|}{$d_{\mathrm{Fe}}=4\ \mathrm{%
\mathring{A}}$} & \multicolumn{1}{|l|}{$5.7$ \cite{HUA+87}} \\ \hline
\multicolumn{1}{|l|}{Fe$_{4}$ / V$_{4}$} & \multicolumn{1}{|l|}{} & 
\multicolumn{1}{|l|}{$1.25$ \cite{LLK+03}} \\ \hline
\multicolumn{1}{|l|}{Fe$_{4}$ / V$_{2}$} & \multicolumn{1}{|l|}{} & 
\multicolumn{1}{|l|}{$0.90$ \cite{LLK+03}} \\ \hline
\multicolumn{1}{|l|}{Ni} & \multicolumn{1}{|l|}{bulk} & \multicolumn{1}{|l|}{%
$2.4$ \cite{HUW02,HUW02a}} \\ \hline
\multicolumn{1}{|l|}{Cu / Co(111)} & \multicolumn{1}{|l|}{} & 
\multicolumn{1}{|l|}{$1.4$ \cite{HWU+03b}} \\ \hline
\multicolumn{1}{|l|}{Co / Cu(001)} & \multicolumn{1}{|l|}{fcc} & 
\multicolumn{1}{|l|}{$3.0$ \cite{SPF+95}} \\ \hline
\multicolumn{1}{|l|}{Co} & \multicolumn{1}{|l|}{fcc, hard} & 
\multicolumn{1}{|l|}{$2.8\pm0.3$ \cite{SPF+95}} \\ \hline
\multicolumn{1}{|l|}{} & \multicolumn{1}{|l|}{easy direction} & 
\multicolumn{1}{|l|}{$1.7\pm0.2$ \cite{SPF+95}} \\ \hline
\multicolumn{1}{|l|}{Fe[001]} & \multicolumn{1}{|l|}{bcc} & 
\multicolumn{1}{|l|}{$0.0(63)$ \cite{KK02}} \\ \hline
\multicolumn{1}{|l|}{Ni[001]} & \multicolumn{1}{|l|}{fcc} & 
\multicolumn{1}{|l|}{$0.(54)$ \cite{KK02}} \\ \hline
\multicolumn{1}{|l|}{Ni[111]} & \multicolumn{1}{|l|}{fcc} & 
\multicolumn{1}{|l|}{$0.(45)$ \cite{KK02}} \\ \hline
\multicolumn{1}{|l|}{Co[0001]} & \multicolumn{1}{|l|}{hcp} & 
\multicolumn{1}{|l|}{$0.0(36)$ \cite{KK02}} \\ \hline
\end{tabular}
\end{table}%

\pagebreak 
%

\begin{table}[hbtp] \centering%
\caption{\label{tab:abc} Third order Taylor series expansion coeffiecients of
the total energy in case of Co$/$Cu$_{n}/$Co. Notice that $\protect b^2+3ac > 0$
and $\protect (3c-a)^2-4b^2$ (not given here) are the smallest in magnitude for
$\protect n=30$, see Eq.\ (\ref{eq:dt-pw}).}%
\begin{tabular}{|c|c|c|l|}
\hline
$n$ & $a$ $(\mathrm{meV})$ & $b$ $(\mathrm{meV})$ & $c$ $(\mathrm{meV})$ \\ 
\hline
$20$ & \multicolumn{1}{|r|}{$0.18482$} & \multicolumn{1}{|r|}{$1.21031$} & 
\multicolumn{1}{|r|}{$1.34444$} \\ \hline
$21$ & \multicolumn{1}{|r|}{$0.10742$} & \multicolumn{1}{|r|}{$0.38908$} & 
\multicolumn{1}{|r|}{$0.57940$} \\ \hline
$22$ & \multicolumn{1}{|r|}{$-0.01466$} & \multicolumn{1}{|r|}{$-0.46855$} & 
\multicolumn{1}{|r|}{$-0.34401$} \\ \hline
$23$ & \multicolumn{1}{|r|}{$0.05495$} & \multicolumn{1}{|r|}{$0.33396$} & 
\multicolumn{1}{|r|}{$0.38826$} \\ \hline
$24$ & \multicolumn{1}{|r|}{$-0.01887$} & \multicolumn{1}{|r|}{$-0.21263$} & 
\multicolumn{1}{|r|}{$-0.19021$} \\ \hline
$25$ & \multicolumn{1}{|r|}{$0.09246$} & \multicolumn{1}{|r|}{$0.30579$} & 
\multicolumn{1}{|r|}{$0.49206$} \\ \hline
$26$ & \multicolumn{1}{|r|}{$0.06293$} & \multicolumn{1}{|r|}{$0.65673$} & 
\multicolumn{1}{|r|}{$0.56195$} \\ \hline
$27$ & \multicolumn{1}{|r|}{$-0.05575$} & \multicolumn{1}{|r|}{$-1.03111$} & 
\multicolumn{1}{|r|}{$-0.78382$} \\ \hline
$28$ & \multicolumn{1}{|r|}{$0.21518$} & \multicolumn{1}{|r|}{$1.54457$} & 
\multicolumn{1}{|r|}{$1.64362$} \\ \hline
$29$ & \multicolumn{1}{|r|}{$-0.01959$} & \multicolumn{1}{|r|}{$-0.25292$} & 
\multicolumn{1}{|r|}{$-0.20745$} \\ \hline
$30$ & \multicolumn{1}{|r|}{$-0.02395$} & \multicolumn{1}{|r|}{$-0.17970$} & 
\multicolumn{1}{|r|}{$-0.12111$} \\ \hline
$31$ & \multicolumn{1}{|r|}{$0.20914$} & \multicolumn{1}{|r|}{$1.64070$} & 
\multicolumn{1}{|r|}{$1.60557$} \\ \hline
$32$ & \multicolumn{1}{|r|}{$-0.24502$} & \multicolumn{1}{|r|}{$-1.92659$} & 
\multicolumn{1}{|r|}{$-1.93901$} \\ \hline
$33$ & \multicolumn{1}{|r|}{$0.17413$} & \multicolumn{1}{|r|}{$1.22408$} & 
\multicolumn{1}{|r|}{$1.28963$} \\ \hline
\end{tabular}
\end{table}%

\pagebreak 
%

\begin{table}[hbtp] \centering%
\caption{\label{tab:taumin} Switching times  (ns) as obtained by using
the third order Taylor series expansion coeffiecients of the total energy,
see Tab.\ \ref{tab:abc}  for  Co$/$Cu$_{n}/$Co$_{m}$. FS (P or AP)
denotes the final magnetic configuration by assuming that for the for the thick
Co slab the direction of the moment  is $\protect n^{(I)}_{\mathrm{z}} = +1$.}%
\begin{tabular}{|c|c|c|c|l|l|l|l|l|}
\hline
$n$ & $m$ & $n_{\mathrm{z}}^{i}$ & $n_{\mathrm{z}}^{f}$ & $FS$ & $\tau_{%
\mathrm{\min}}$ & $\tau(\alpha_{\mathrm{G}}^{\mathrm{bulk}})$ \cite{TBB02} & 
$\tau(\alpha_{\mathrm{G}}^{\mathrm{Co/Cu/Co}})$ \cite{Ber01} & $\tau(\alpha_{%
\mathrm{G}}^{\mathrm{scaled}})\text{ \cite{MRK+99}}$ \\ \hline
$20$ & $13$ & $-1$ & $+1$ & P & $0.01340$ & \multicolumn{1}{|r|}{$1.33999$}
& \multicolumn{1}{|r|}{$0.54039$} & \multicolumn{1}{|c|}{$%
\allowbreak0.108\,24$} \\ \hline
$21$ & $12$ & $-1$ & $+1$ & P & $0.02506$ & \multicolumn{1}{|r|}{$2.50591$}
& \multicolumn{1}{|r|}{$1.01058$} & \multicolumn{1}{|c|}{$%
\allowbreak0.186\,97$} \\ \hline
$22$ & $14$ & $+1$ & $-1$ & AP & $0.18977$ & \multicolumn{1}{|r|}{$18.97722$}
& \multicolumn{1}{|r|}{$7.65309$} & \multicolumn{1}{|c|}{$\allowbreak
1.\,\allowbreak649\,90$} \\ \hline
$23$ & $13$ & $-1$ & $+1$ & P & $0.04415$ & \multicolumn{1}{|r|}{$4.41567$}
& \multicolumn{1}{|r|}{$1.78074$} & \multicolumn{1}{|c|}{$%
\allowbreak0.356\,67$} \\ \hline
$24$ & $12$ & $+1$ & $-1$ & AP & $0.13645$ & \multicolumn{1}{|r|}{$13.64524$}
& \multicolumn{1}{|r|}{$5.50282$} & \multicolumn{1}{|c|}{$\allowbreak
1.\,\allowbreak018\,10$} \\ \hline
$25$ & $14$ & $-1$ & $+1$ & P = AP & $0.02841$ & \multicolumn{1}{|r|}{$%
2.84111$} & \multicolumn{1}{|r|}{$1.14576$} & \multicolumn{1}{|c|}{$%
\allowbreak0.247\,01$} \\ \hline
$26$ & $13$ & $-1$ & $+1$ & P $\simeq$ GS & $0.05414$ & \multicolumn{1}{|r|}{%
$5.41415$} & \multicolumn{1}{|r|}{$2.18341$} & \multicolumn{1}{|c|}{$%
\allowbreak0.437\,33$} \\ \hline
$27$ & $12$ & $+1$ & $-1$ & AP & $0.06667$ & \multicolumn{1}{|r|}{$6.66771$}
& \multicolumn{1}{|r|}{$2.68894$} & \multicolumn{1}{|c|}{$%
\allowbreak0.497\,49$} \\ \hline
$28$ & $11$ & $-1$ & $+1$ & P & $0.01154$ & \multicolumn{1}{|r|}{$1.15445$}
& \multicolumn{1}{|r|}{$0.46556$} & \multicolumn{1}{|c|}{$0.07902$} \\ \hline
$29$ & $13$ & $+1$ & $-1$ & AP & $0.16889$ & \multicolumn{1}{|r|}{$16.88991$}
& \multicolumn{1}{|r|}{$6.81132$} & \multicolumn{1}{|c|}{$\allowbreak
1.\,\allowbreak364\,30$} \\ \hline
$30$ & $12$ & $-1$ & $+1$ & P (AP) & $0.68836$ & \multicolumn{1}{|r|}{$%
68.83831$} & \multicolumn{1}{|r|}{$27.76096$} & \multicolumn{1}{|c|}{$%
\allowbreak5.\,\allowbreak136\,10$} \\ \hline
$31$ & $14$ & $-1$ & $+1$ & P & $0.01358$ & \multicolumn{1}{|r|}{$1.35829$}
& \multicolumn{1}{|r|}{$0.54777$} & \multicolumn{1}{|c|}{$%
\allowbreak0.118\,09$} \\ \hline
$32$ & $13$ & $+1$ & $-1$ & AP & $0.01064$ & \multicolumn{1}{|r|}{$1.06400$}
& \multicolumn{1}{|r|}{$0.42909$} & \multicolumn{1}{|c|}{$\allowbreak
0.08\,\allowbreak594\,$} \\ \hline
$33$ & $12$ & $-1$ & $+1$ & P & $0.01498$ & \multicolumn{1}{|r|}{$1.49796$}
& \multicolumn{1}{|r|}{$0.60409$} & \multicolumn{1}{|c|}{$%
\allowbreak0.111\,76$} \\ \hline
\end{tabular}
\end{table}%

\pagebreak 
%

\begin{table}[tbp] \centering%
\caption{\label{tab:cross-sect} Cross section of multilayer pillar sequence
Co$/$Cu$/$Co used in experiments.}%
\begin{tabular}{|l|l|l|}
\hline
\textbf{multilayer pillar sequence} & \textbf{cross section} & $A_{0}\
\left( \mathrm{nm}^{2}\right) $ \\ \hline
Co($100$ $\mathrm{nm}$)$/$Cu($4$ $\mathrm{nm}$)$/$Co($d_{\mathrm{Co}}$), & $%
5\div10$ $\mathrm{nm}$ diameter & \multicolumn{1}{|r|}{$\left( 1\div4\right)
\cdot19.63$} \\ \hline
$\qquad d_{\mathrm{Co}}=2,4,7,10$ $\mathrm{nm}$ \cite{MRK+99} &  &  \\ \hline
Co($100$ $\mathrm{\mathring{A}}$)$/$Cu($60$ $\mathrm{\mathring{A}}$)$/$Co($%
25 $ $\mathrm{\mathring{A}}$) \cite{KAB+00} & $130\pm30$ $\mathrm{nm}$
diameter & \multicolumn{1}{|r|}{$7853.98\div20106.19$} \\ \hline
Co($40$ $\mathrm{nm}$)$/$Cu($6$ $\mathrm{nm}$)$/$Co($3$ $\mathrm{nm}$) \cite%
{MAS+02} & $\sim50\times50$ $\mathrm{nm}$ (sample 1) & \multicolumn{1}{|r|}{$%
2500.00$} \\ \hline
& $\sim130\times60$ $\mathrm{nm}$ (sample 2) & \multicolumn{1}{|r|}{$7800.00$%
} \\ \hline
Co($t_{\mathrm{Fixed}}$ $\mathrm{nm}$)$/$Cu($d_{\mathrm{Cu}}$ $\mathrm{nm}$)$%
/$Co($t_{\mathrm{Free}}$ $\mathrm{nm}$) & $\leq100$ $\mathrm{nm}$ diameter & 
$7853.98$ \\ \hline
$\qquad t_{\mathrm{Fixed}}\geq4t_{\mathrm{Free}}$ \cite{AEM+02} &  & 
\multicolumn{1}{|r|}{} \\ \hline
Co($30$ $\mathrm{nm}$)$/$Cu($10$ $\mathrm{nm}$)$/$Co($10$ $\mathrm{nm}$) 
\cite{WHG+02} & $\sim60\div80$ $\mathrm{nm}$ diameter & \multicolumn{1}{|r|}{%
$2827.43\div5026.55$} \\ \hline
Co($15$ $\mathrm{nm}$)$/$Cu($10$ $\mathrm{nm}$)$/$Co($2.5$ $\mathrm{nm}$) 
\cite{GCJ+03} & $200\times600$ $\mathrm{nm}^{2}$ & \multicolumn{1}{|r|}{$%
120000.00$} \\ \hline
Co($10$ $\mathrm{nm}$)$/$Cu($10$ $\mathrm{nm}$)$/$Co($30$ $\mathrm{nm}$) 
\cite{FTG+03} & $\sim40$ $\mathrm{nm}$ diameter & \multicolumn{1}{|r|}{$%
1256.64$} \\ \hline
Co($3$ $\mathrm{nm}$)$/$Cu($10$ $\mathrm{nm}$)$/$Co($12$ $\mathrm{nm}$) \cite%
{OKM+03} & $\sim100$ $\mathrm{nm}$ diameter & \multicolumn{1}{|r|}{$7853.98$}
\\ \hline
Co($10$ $\mathrm{nm}$)$/$Cu($10$ $\mathrm{nm}$)$/$Co($30$ $\mathrm{nm}$) 
\cite{Weg03} & $\sim40$ $\mathrm{nm}$ diameter & \multicolumn{1}{|r|}{$%
1256.64$} \\ \hline
Co($3$ $\mathrm{nm}$)$/$Cu($10$ $\mathrm{nm}$)$/$Co($12$ $\mathrm{nm}$) \cite%
{SMK+03} & $0.05\times0.10$ $\mathrm{\mu m}^{2}$ & \multicolumn{1}{|r|}{$%
5000.00$} \\ \hline
& $0.05\times0.20$ $\mathrm{\mu m}^{2}$ & \multicolumn{1}{|r|}{$10000.00$}
\\ \hline
& $0.07\times0.14$ $\mathrm{\mu m}^{2}$ & \multicolumn{1}{|r|}{$9800.00$} \\ 
\hline
& $0.08\times0.16$ $\mathrm{\mu m}^{2}$ & \multicolumn{1}{|r|}{$12800.00$}
\\ \hline
Co($3$ $\mathrm{nm}$)$/$Cu($10$ $\mathrm{nm}$)$/$Co($12$ $\mathrm{nm}$) \cite%
{TSR+04} & $50\div200$ $\mathrm{nm}$ circumference & \multicolumn{1}{|r|}{$%
198.94\div3183.10$} \\ \hline
Co$/$Cu$/$Co($30$ $\mathrm{\mathring{A}}$) \cite{KKS04} & $0.05\times0.10$ $%
\mathrm{\mu m}^{2}$ & \multicolumn{1}{|r|}{$5000.00$} \\ \hline
\end{tabular}
\end{table}%
\end{center}

\begin{figure}[tbph]
\begin{center}
\includegraphics[width=0.85\textwidth]{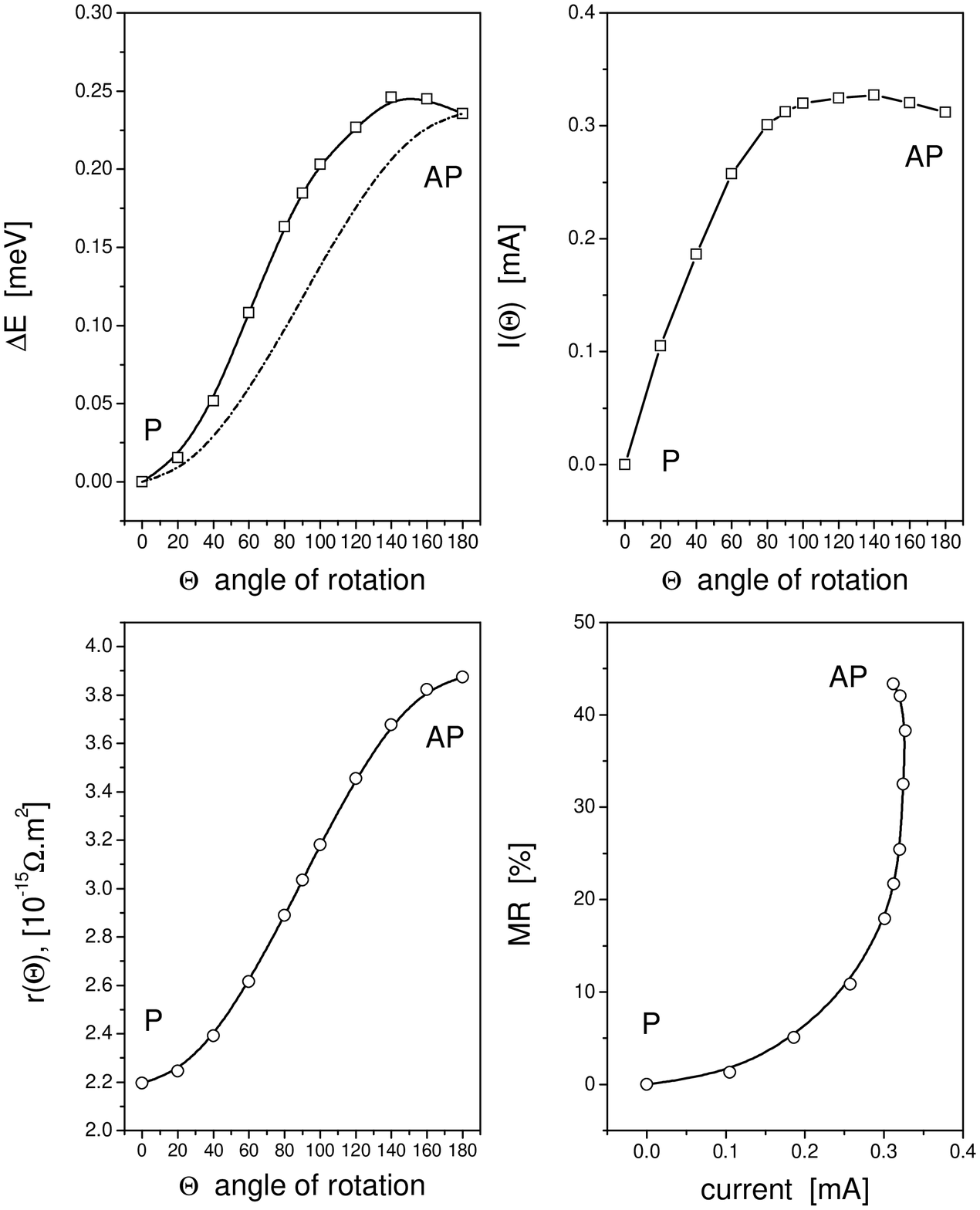}
\end{center}
\caption{ Co/Cu$_{20}$/Co, spacer thickness: 34.65 \AA . Left column:
twisting energy and sheet resistance as a function of the rotation angle $%
\Theta $. The dashed-dotted line refers to the first order approximation for
the twisting energy. Right column: current as a function of the rotation
angle $\Theta $ (top) and magnetoresistance as a function of the current
(bottom), $\protect\sqrt{\left\langle A_{0}\right\rangle _{\mathrm{SI}%
}/\left\langle \protect\tau _{\min }\right\rangle _{\mathrm{SI}}}=1$, see
Eq.(\protect\ref{e-final}). Solid lines serve as guidance for the eye.}
\label{fig:fig1}
\end{figure}
\begin{figure}[tbph]
\centering \includegraphics[width=0.85\textwidth]{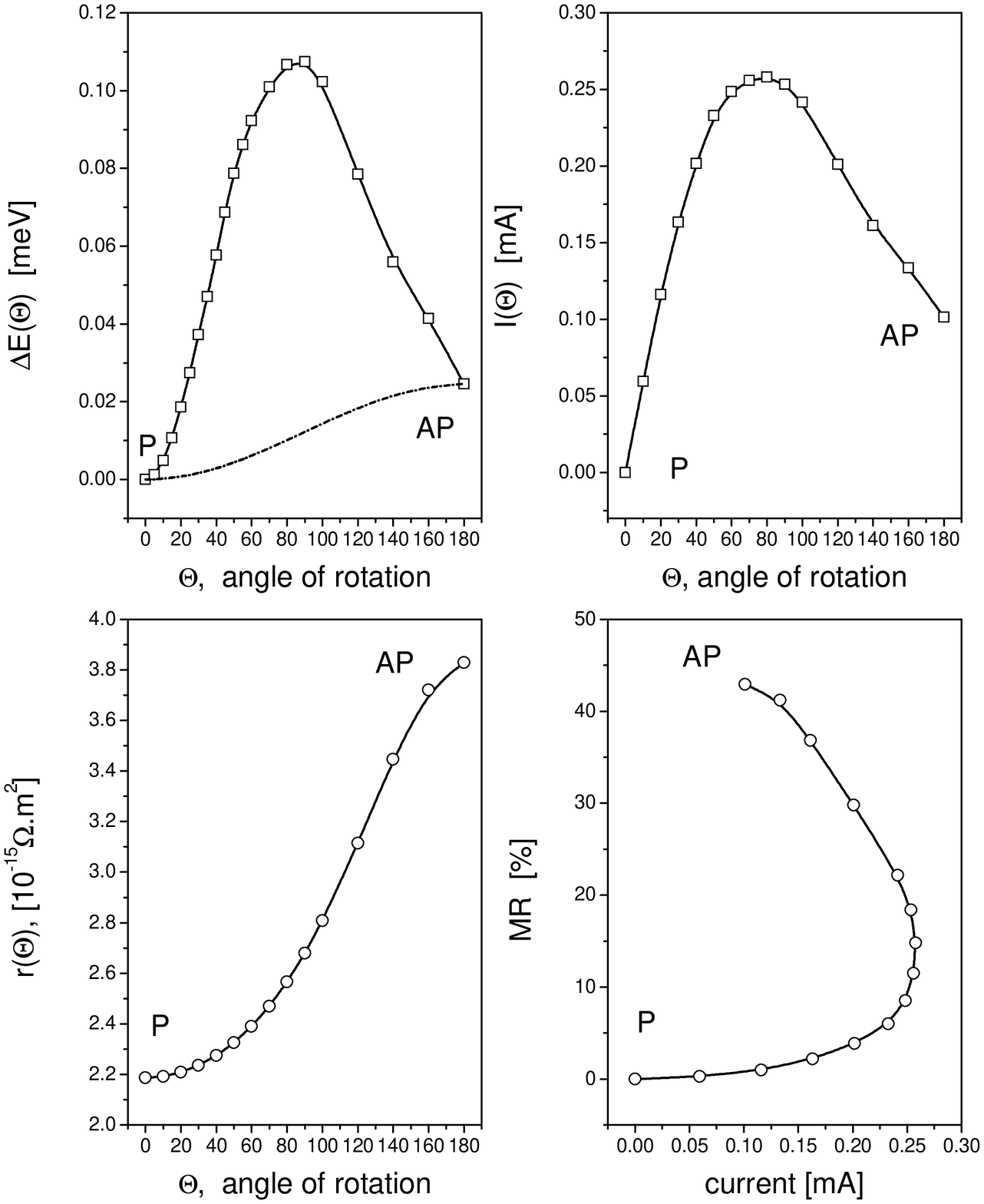}
\caption{ Co/Cu$_{21}$/Co, spacer thickness: 36.39 \AA . Left column:
twisting energy and sheet resistance as a function of the rotation angle $%
\Theta $. The dashed-dotted line refers to the first order approximation for
the twisting energy. Right column: current as a function of the rotation
angle $\Theta $ (top) and magnetoresistance as a function of the current
(bottom), $\protect\sqrt{\left\langle A_{0}\right\rangle _{\mathrm{SI}%
}/\left\langle \protect\tau _{\min }\right\rangle _{\mathrm{SI}}}=1$, see
Eq.(\protect\ref{e-final}). Solid lines serve as guidance for the eye.}
\end{figure}
\begin{figure}[tbph]
\centering \includegraphics[width=0.85\textwidth]{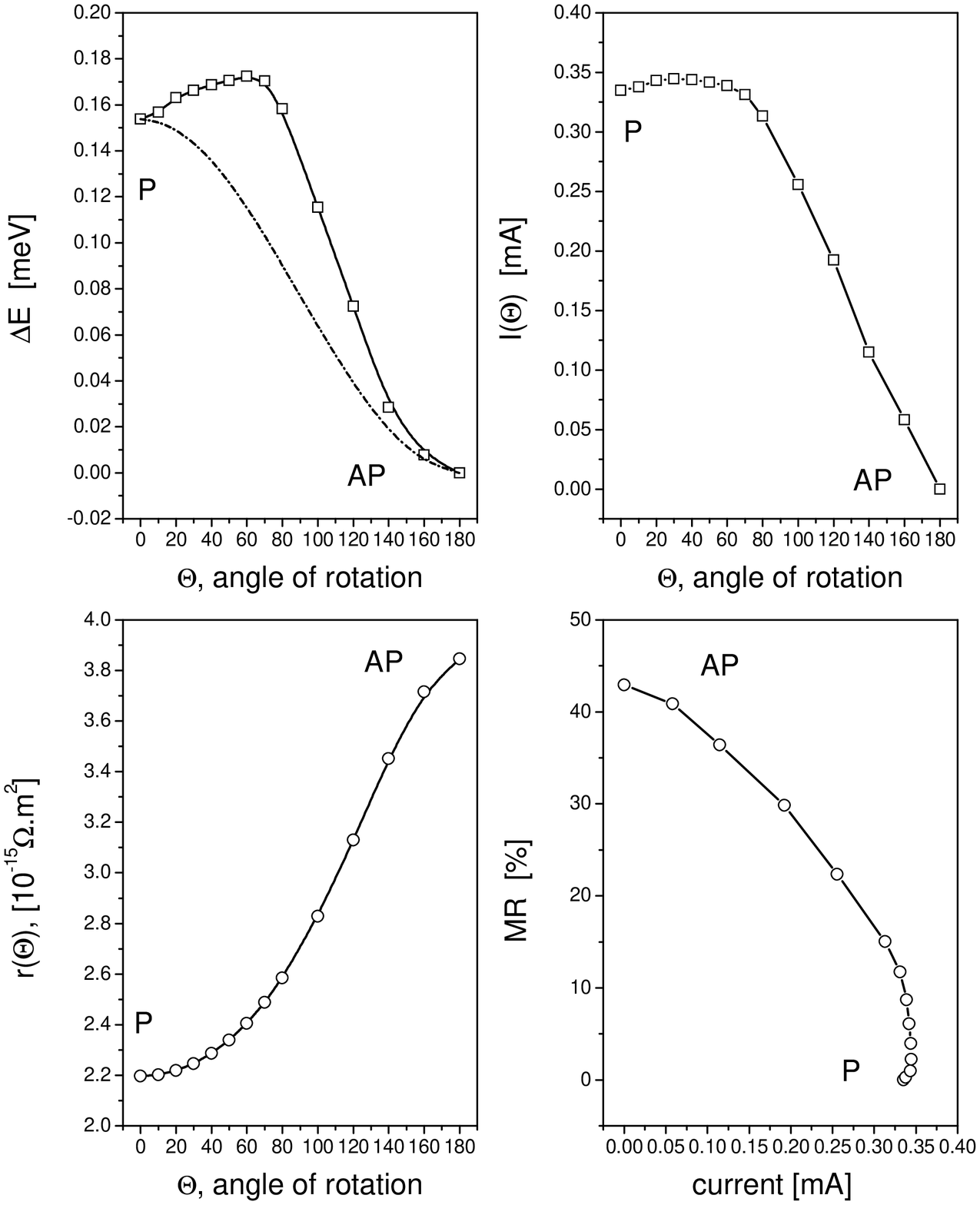}
\caption{ Co/Cu$_{22}$/Co, spacer thickness: 38.12 \AA . Left column:
twisting energy and sheet resistance as a function of the rotation angle $%
\Theta $. The dashed-dotted line refers to the first order approximation for
the twisting energy. Right column: current as a function of the rotation
angle $\Theta $ (top) and magnetoresistance as a function of the current
(bottom), $\protect\sqrt{\left\langle A_{0}\right\rangle _{\mathrm{SI}%
}/\left\langle \protect\tau _{\min }\right\rangle _{\mathrm{SI}}}=1$, see
Eq.(\protect\ref{e-final}). Solid lines serve as guidance for the eye.}
\end{figure}
\begin{figure}[tbph]
\centering \includegraphics[width=0.85\textwidth]{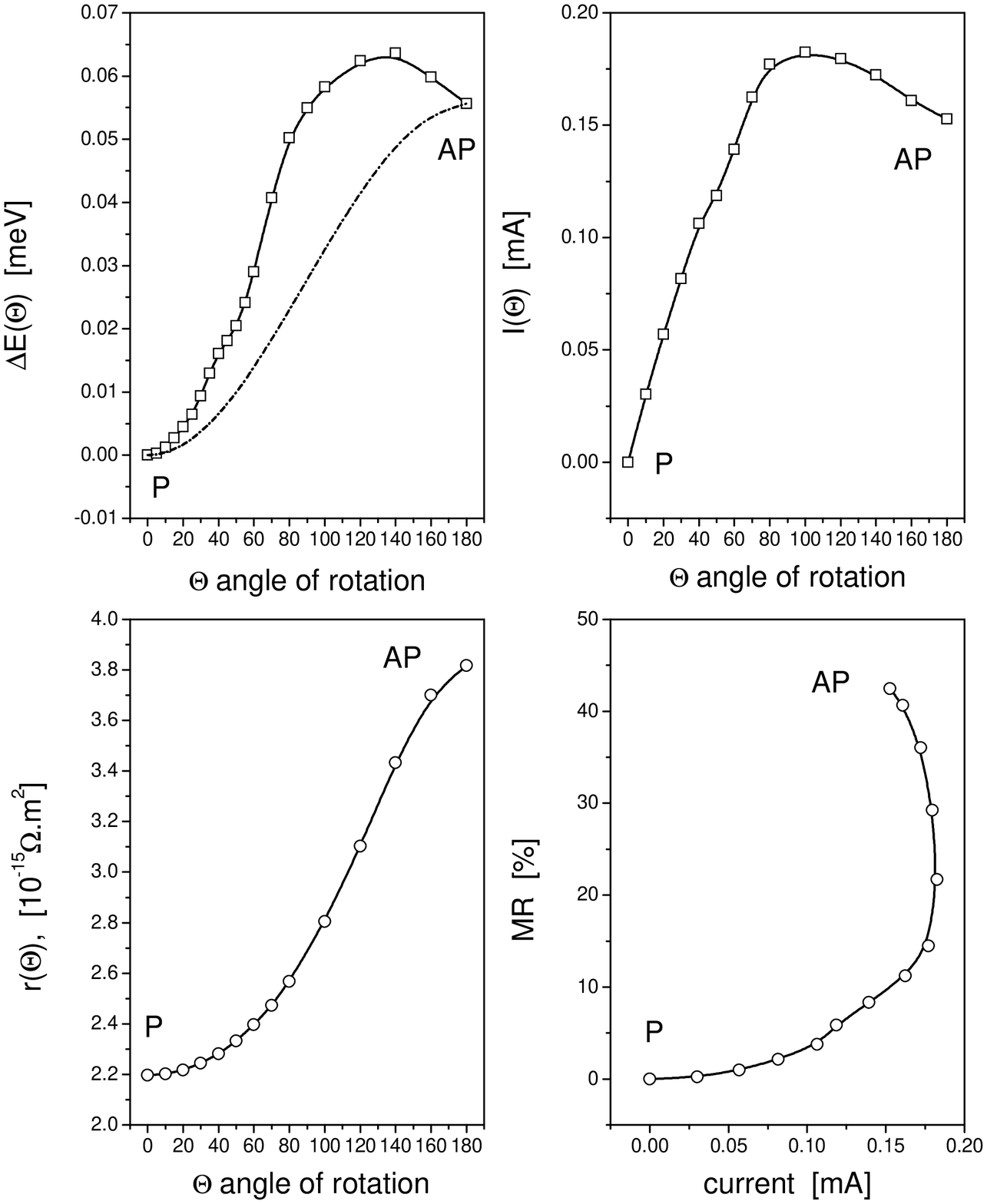}
\caption{ Co/Cu$_{23}$/Co, spacer thickness: 39.85 \AA . Left column:
twisting energy and sheet resistance as a function of the rotation angle $%
\Theta $. The dashed-dotted line refers to the first order approximation for
the twisting energy. Right column: current as a function of the rotation
angle $\Theta $ (top) and magnetoresistance as a function of the current
(bottom), $\protect\sqrt{\left\langle A_{0}\right\rangle _{\mathrm{SI}%
}/\left\langle \protect\tau _{\min }\right\rangle _{\mathrm{SI}}}=1$, see
Eq.(\protect\ref{e-final}). Solid lines serve as guidance for the eye.}
\end{figure}
\begin{figure}[tbph]
\centering \includegraphics[width=0.85\textwidth]{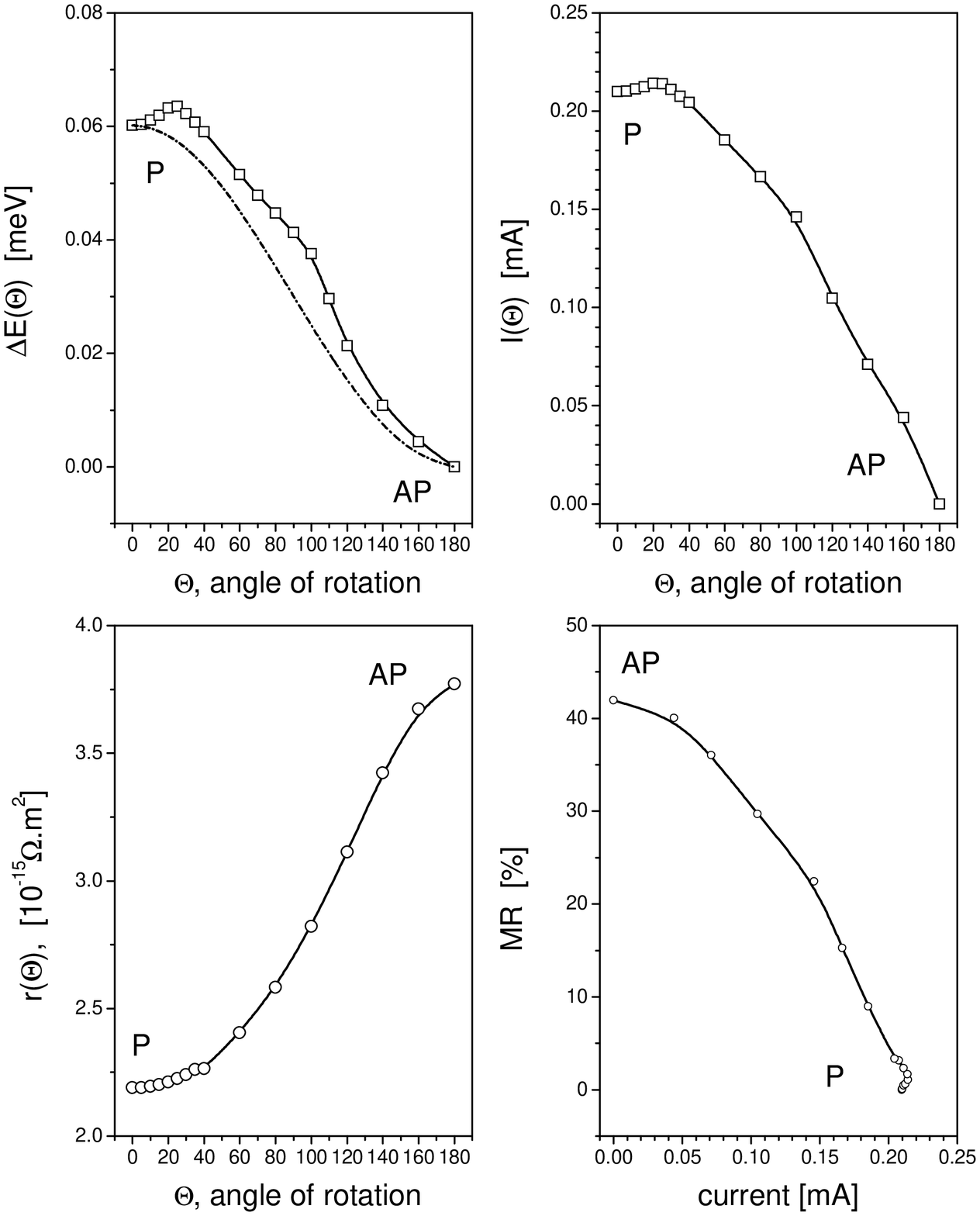}
\caption{ Co/Cu$_{24}$/Co, spacer thickness: 41.58 \AA . Left column:
twisting energy and sheet resistance as a function of the rotation angle $%
\Theta $. The dashed-dotted line refers to the first order approximation for
the twisting energy. Right column: current as a function of the rotation
angle $\Theta $ (top) and magnetoresistance as a function of the current
(bottom), $\protect\sqrt{\left\langle A_{0}\right\rangle _{\mathrm{SI}%
}/\left\langle \protect\tau _{\min }\right\rangle _{\mathrm{SI}}}=1$, see
Eq.(\protect\ref{e-final}). Solid lines serve as guidance for the eye.}
\end{figure}
\begin{figure}[tbph]
\centering \includegraphics[width=0.85\textwidth]{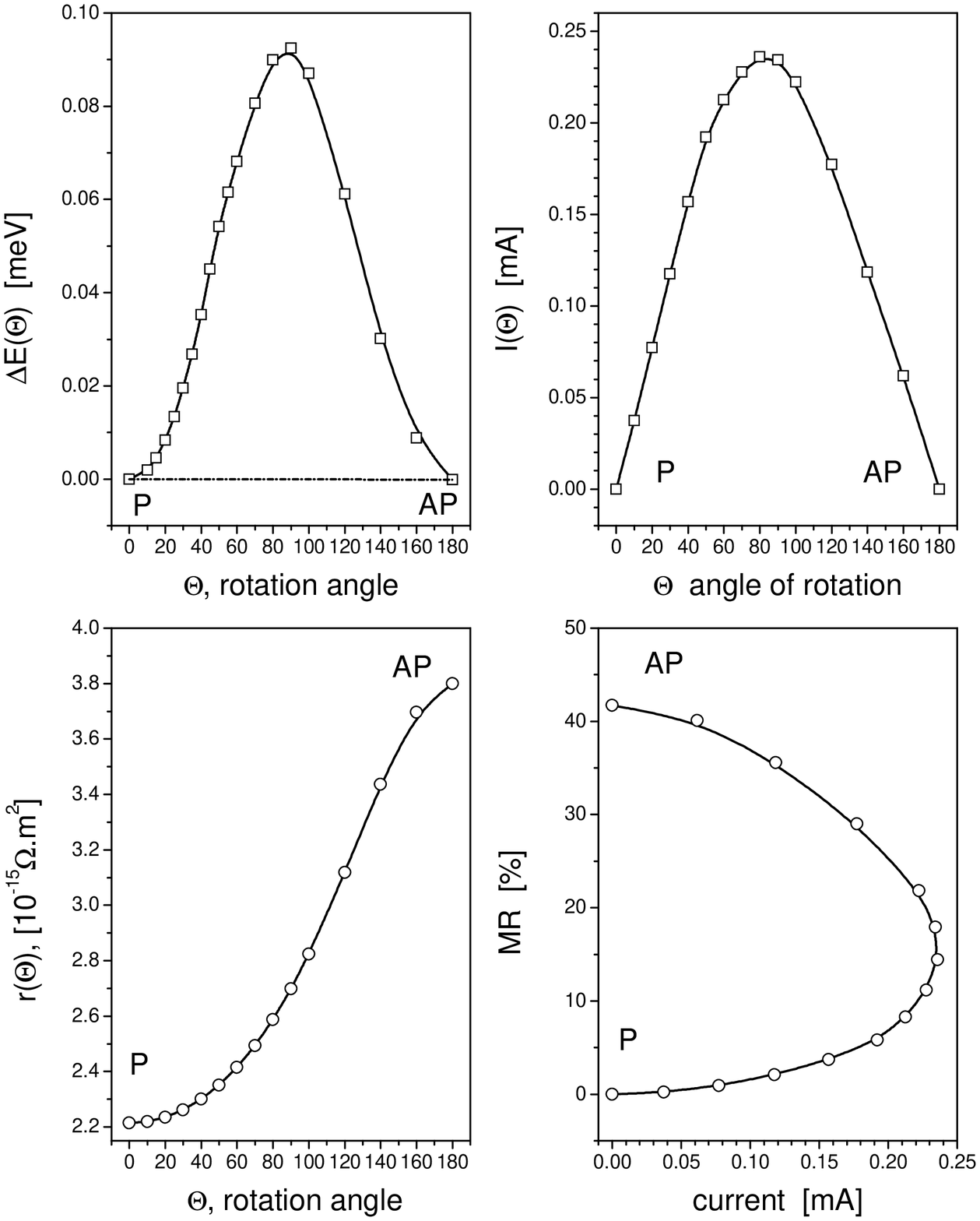}
\caption{ Co/Cu$_{25}$/Co, spacer thickness: 43.32 \AA . Left column:
twisting energy and sheet resistance as a function of the rotation angle $%
\Theta $. The dashed-dotted line refers to the first order approximation for
the twisting energy. Right column: current as a function of the rotation
angle $\Theta $ (top) and magnetoresistance as a function of the current
(bottom), $\protect\sqrt{\left\langle A_{0}\right\rangle _{\mathrm{SI}%
}/\left\langle \protect\tau _{\min }\right\rangle _{\mathrm{SI}}}=1$, see
Eq.(\protect\ref{e-final}). Solid lines serve as guidance for the eye.}
\end{figure}
\begin{figure}[tbph]
\centering \includegraphics[width=0.85\textwidth]{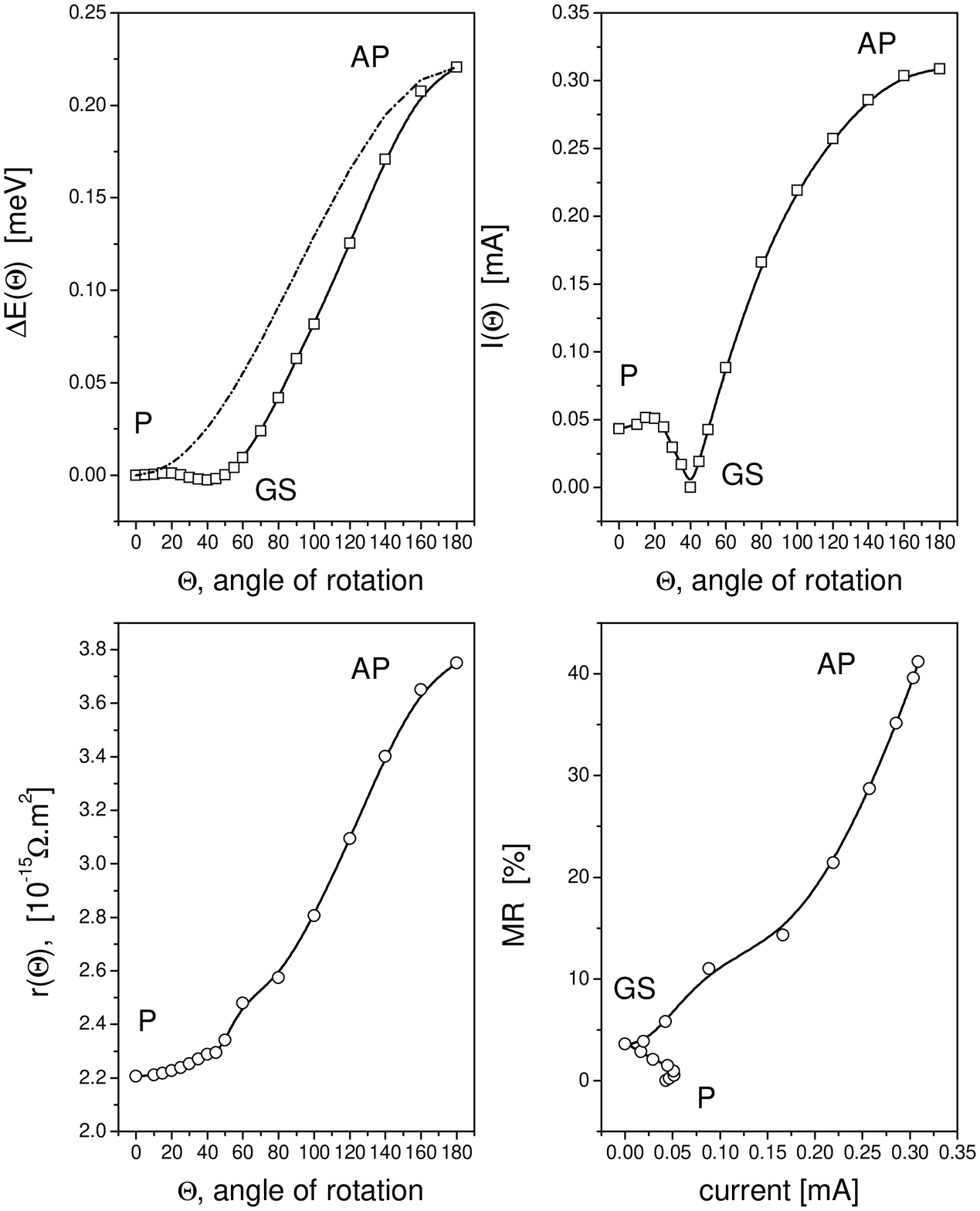}
\caption{ Co/Cu$_{26}$/Co, spacer thickness: 45.05 \AA . Left column:
twisting energy and sheet resistance as a function of the rotation angle $%
\Theta $. The dashed-dotted line refers to the first order approximation for
the twisting energy. Right column: current as a function of the rotation
angle $\Theta $ (top) and magnetoresistance as a function of the current
(bottom), $\protect\sqrt{\left\langle A_{0}\right\rangle _{\mathrm{SI}%
}/\left\langle \protect\tau _{\min }\right\rangle _{\mathrm{SI}}}=1$, see
Eq.(\protect\ref{e-final}). Solid lines serve as guidance for the eye.}
\end{figure}
\begin{figure}[tbph]
\centering \includegraphics[width=0.85\textwidth,clip]{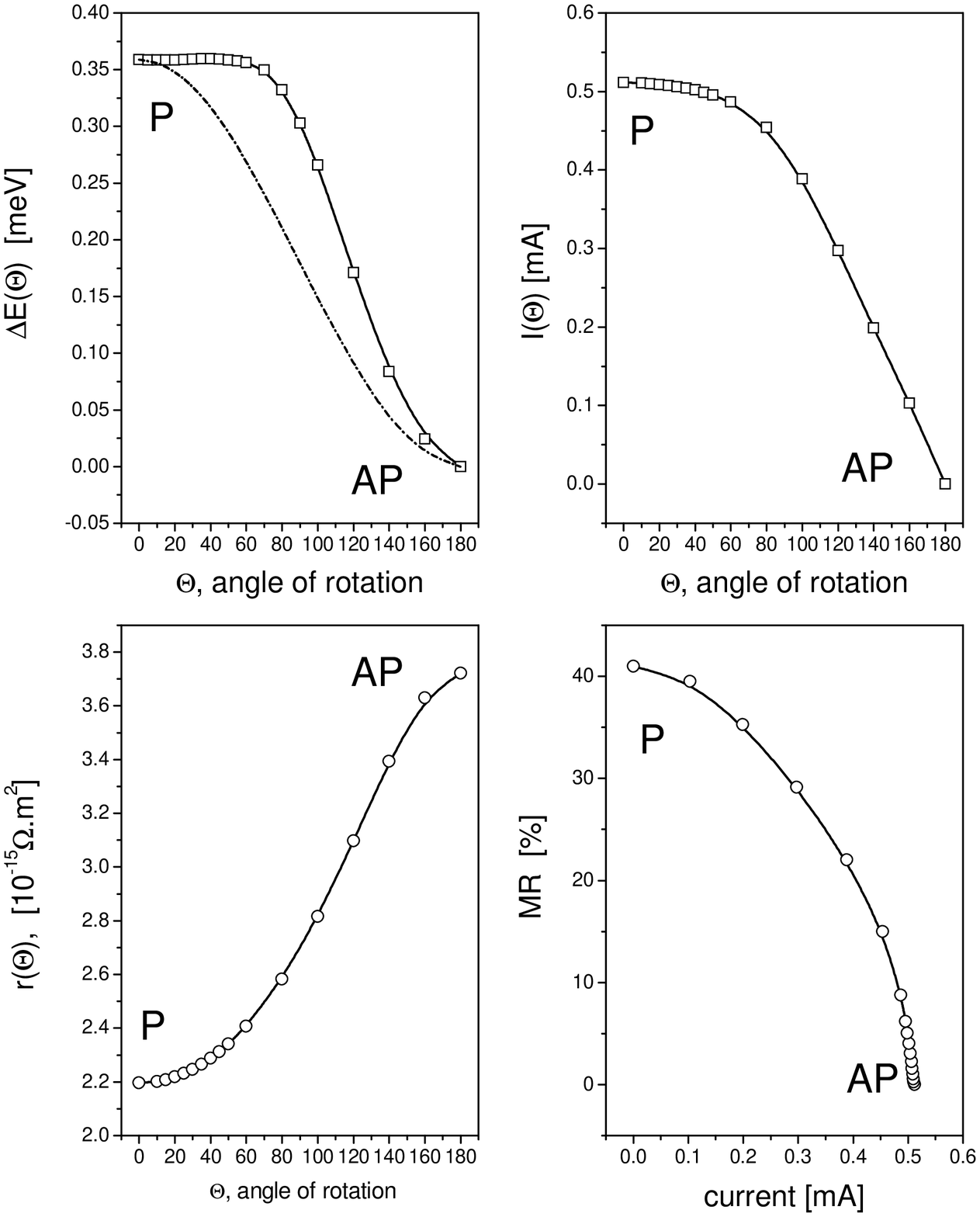}
\caption{ Co/Cu$_{27}$/Co, spacer thickness: 46.78 \AA . Left column:
twisting energy and sheet resistance as a function of the rotation angle $%
\Theta $. The dashed-dotted line refers to the first order approximation for
the twisting energy. Right column: current as a function of the rotation
angle $\Theta $ (top) and magnetoresistance as a function of the current
(bottom), $\protect\sqrt{\left\langle A_{0}\right\rangle _{\mathrm{SI}%
}/\left\langle \protect\tau _{\min }\right\rangle _{\mathrm{SI}}}=1$, see
Eq.(\protect\ref{e-final}). Solid lines serve as guidance for the eye.}
\end{figure}
\begin{figure}[tbph]
\centering \includegraphics[width=0.85\textwidth]{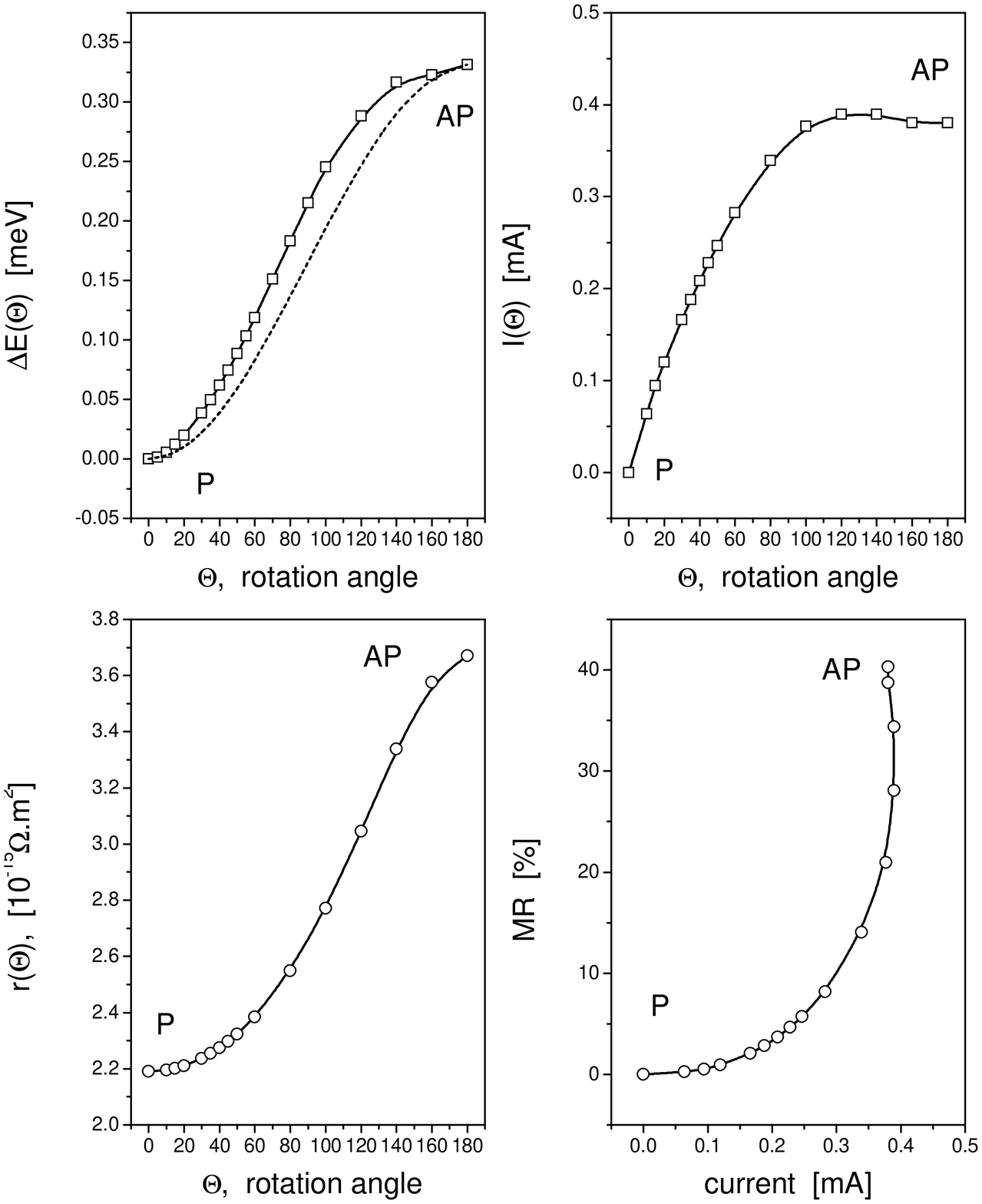}
\caption{ Co/Cu$_{28}$/Co, spacer thickness: 48.51 \AA . Left column:
twisting energy and sheet resistance as a function of the rotation angle $%
\Theta $. The dashed-dotted line refers to the first order approximation for
the twisting energy. Right column: current as a function of the rotation
angle $\Theta $ (top) and magnetoresistance as a function of the current
(bottom), $\protect\sqrt{\left\langle A_{0}\right\rangle _{\mathrm{SI}%
}/\left\langle \protect\tau _{\min }\right\rangle _{\mathrm{SI}}}=1$, see
Eq.(\protect\ref{e-final}). Solid lines serve as guidance for the eye.}
\end{figure}
\begin{figure}[tbph]
\centering \includegraphics[width=0.85\textwidth]{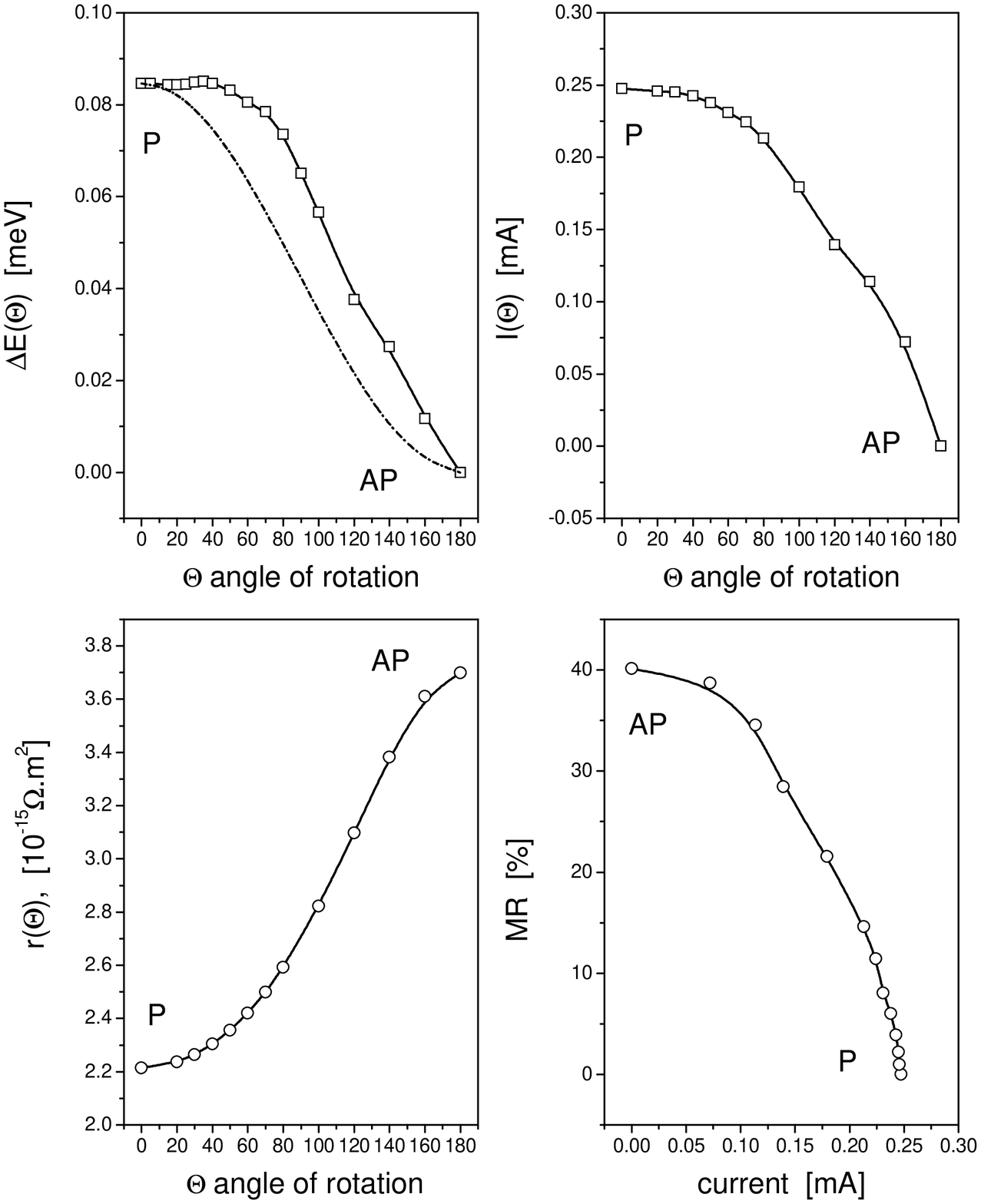}
\caption{ Co/Cu$_{29}$/Co, spacer thickness: 50.24 \AA . Left column:
twisting energy and sheet resistance as a function of the rotation angle $%
\Theta $. The dashed-dotted line refers to the first order approximation for
the twisting energy. Right column: current as a function of the rotation
angle $\Theta $ (top) and magnetoresistance as a function of the current
(bottom), $\protect\sqrt{\left\langle A_{0}\right\rangle _{\mathrm{SI}%
}/\left\langle \protect\tau _{\min }\right\rangle _{\mathrm{SI}}}=1$, see
Eq.(\protect\ref{e-final}). Solid lines serve as guidance for the eye.}
\end{figure}
\begin{figure}[tbph]
\centering \includegraphics[width=0.85\textwidth]{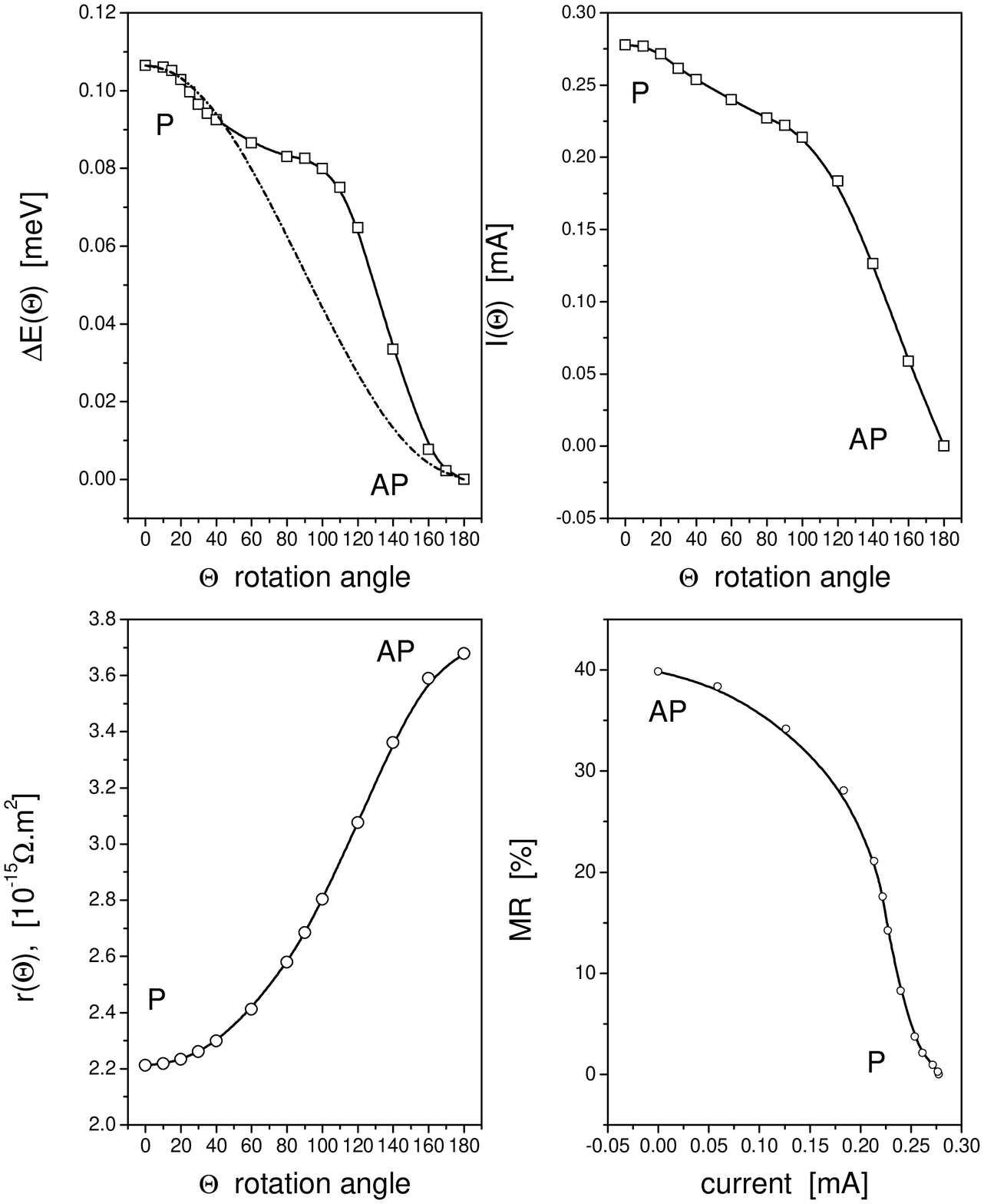}
\caption{ Co/Cu$_{30}$/Co, spacer thickness: 51.98 \AA . Left column:
twisting energy and sheet resistance as a function of the rotation angle $%
\Theta $. The dashed-dotted line refers to the first order approximation for
the twisting energy. Right column: current as a function of the rotation
angle $\Theta $ (top) and magnetoresistance as a function of the current
(bottom), $\protect\sqrt{\left\langle A_{0}\right\rangle _{\mathrm{SI}%
}/\left\langle \protect\tau _{\min }\right\rangle _{\mathrm{SI}}}=1$, see
Eq.(\protect\ref{e-final}). Solid lines serve as guidance for the eye.}
\end{figure}
\begin{figure}[tbph]
\centering \includegraphics[width=0.85\textwidth]{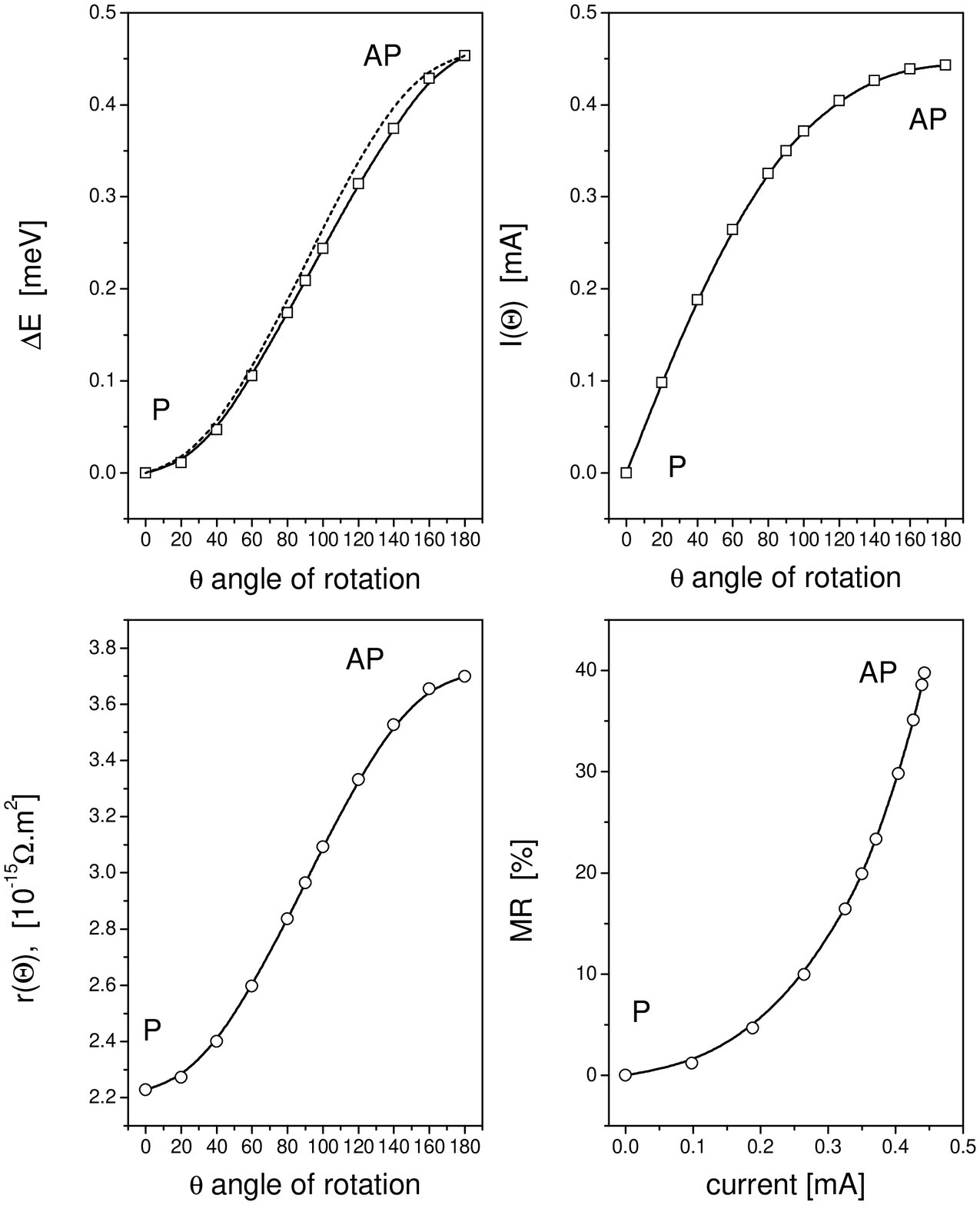}
\caption{ Co/Cu$_{31}$/Co, spacer thickness: 53.71 \AA . Left column:
twisting energy and sheet resistance as a function of the rotation angle $%
\Theta $. The dashed-dotted line refers to the first order approximation for
the twisting energy. Right column: current as a function of the rotation
angle $\Theta $ (top) and magnetoresistance as a function of the current
(bottom), $\protect\sqrt{\left\langle A_{0}\right\rangle _{\mathrm{SI}%
}/\left\langle \protect\tau _{\min }\right\rangle _{\mathrm{SI}}}=1$, see
Eq.(\protect\ref{e-final}). Solid lines serve as guidance for the eye.}
\end{figure}
\begin{figure}[tbph]
\centering \includegraphics[width=0.85\textwidth]{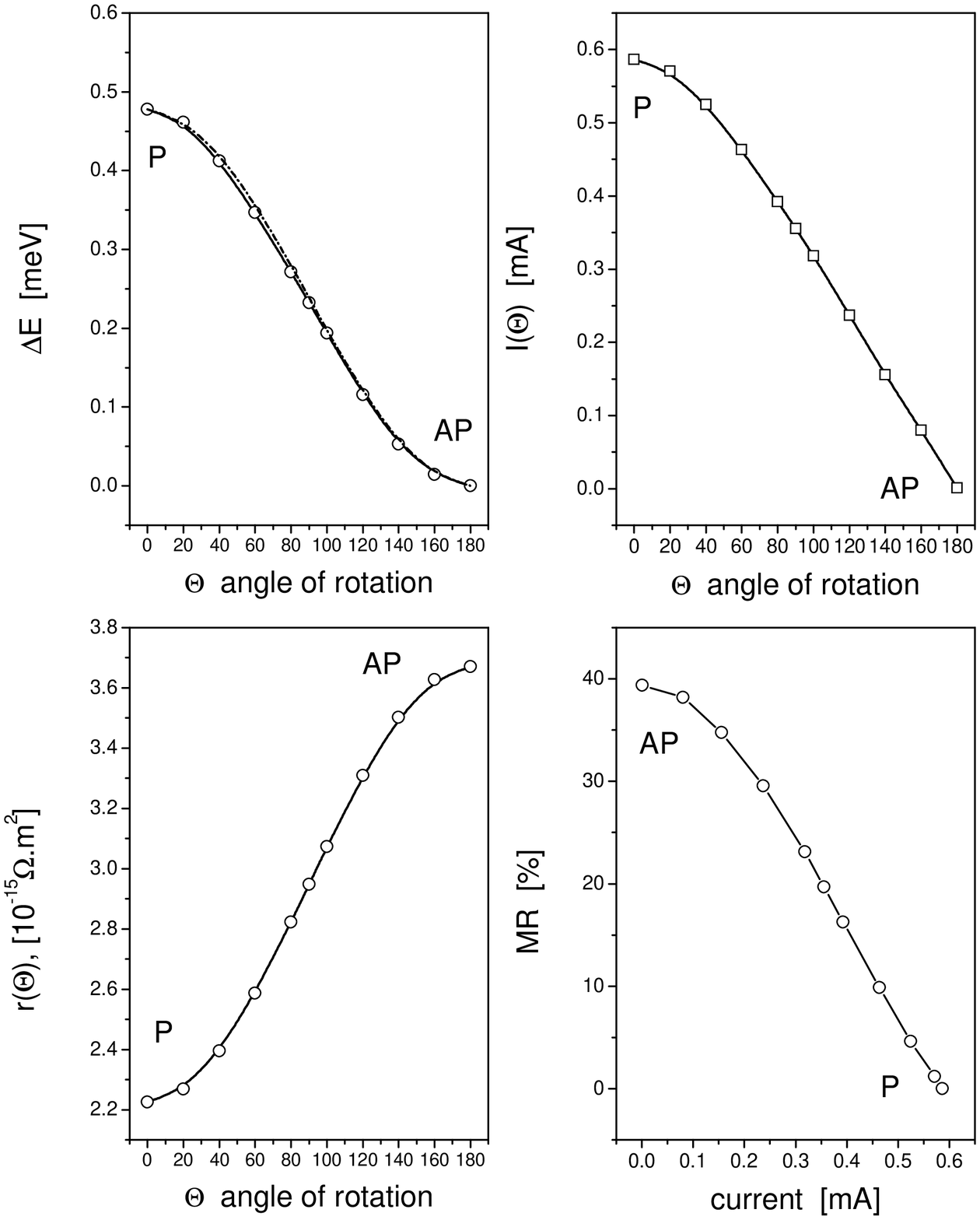}
\caption{ Co/Cu$_{32}$/Co, spacer thickness: 55.45 \AA . Left column:
twisting energy and sheet resistance as a function of the rotation angle $%
\Theta $. The dashed-dotted line refers to the first order approximation for
the twisting energy. Right column: current as a function of the rotation
angle $\Theta $ (top) and magnetoresistance as a function of the current
(bottom), $\protect\sqrt{\left\langle A_{0}\right\rangle _{\mathrm{SI}%
}/\left\langle \protect\tau _{\min }\right\rangle _{\mathrm{SI}}}=1$, see
Eq.(\protect\ref{e-final}). Solid lines serve as guidance for the eye.}
\end{figure}
\begin{figure}[tbph]
\centering \includegraphics[width=0.85\textwidth]{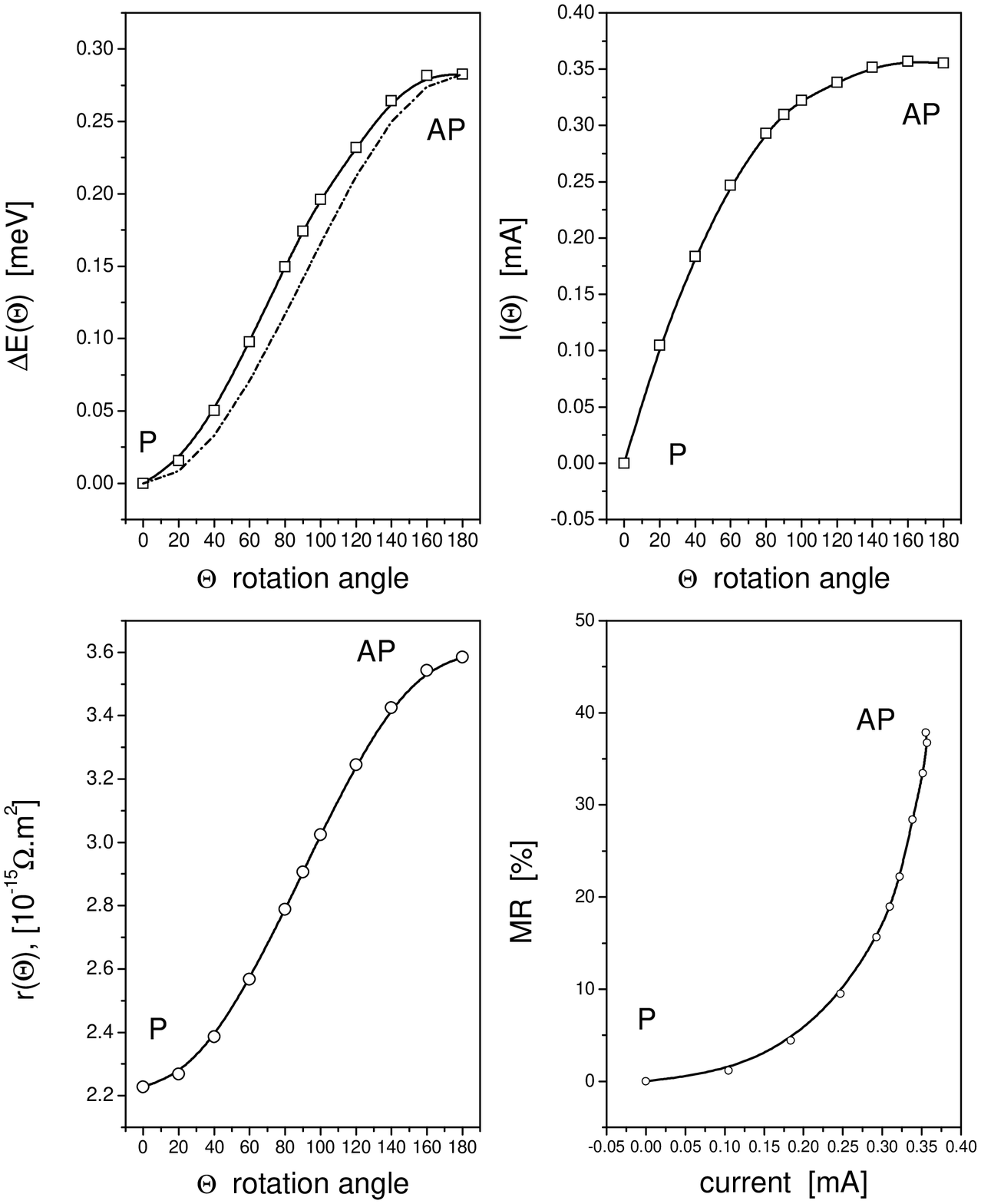}
\caption{ Co/Cu$_{33}$/Co, spacer thickness: 57.18 \AA . Left column:
twisting energy and sheet resistance as a function of the rotation angle $%
\Theta $. The dashed-dotted line refers to the first order approximation for
the twisting energy. Right column: current as a function of the rotation
angle $\Theta $ (top) and magnetoresistance as a function of the current
(bottom), $\protect\sqrt{\left\langle A_{0}\right\rangle _{\mathrm{SI}%
}/\left\langle \protect\tau _{\min }\right\rangle _{\mathrm{SI}}}=1$, see
Eq.(\protect\ref{e-final}). Solid lines serve as guidance for the eye.}
\end{figure}
\begin{figure}[tbph]
\centering \includegraphics[width=0.95\textwidth]{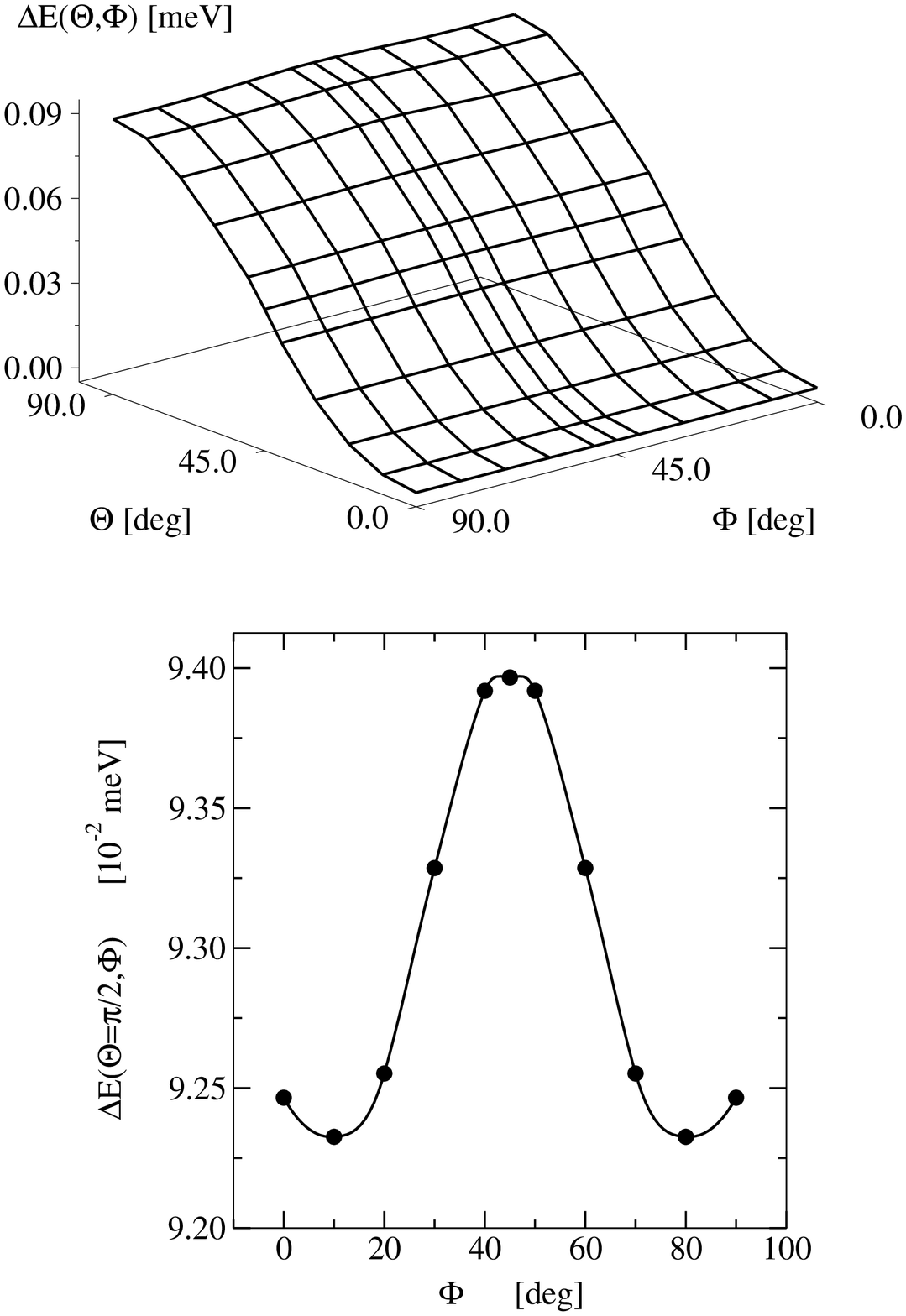}
\caption{ Top: Twisting energy as a function of both rotation angles for the
system with 25 spacer layers of Cu. Bottom: precessional energy at $\Theta
=90^{0}$.}
\label{precession}
\end{figure}

\end{document}